\documentclass[12pt]{article}
\newcommand{\llb}{\left[}
\newcommand{\llm}{\left\{}
\newcommand{\lls}{\left(}

\newcommand{\rrb}{\right]}
\newcommand{\rrm}{\right\}}
\newcommand{\rrs}{\right)}

\newcommand{\tth}[1]{$[$\,\ref{th#1}\,$]$}
\newcommand{\llem}[1]{$[$\,\ref{lem#1}\,$]$}
\newcommand{\ccor}[1]{$[$\,\ref{cor#1}\,$]$}
\newcommand{\ddef}[1]{\ref{def#1}}
\newcommand{\ppn}[1]{$[$\,\ref{ppn#1}\,$]$}
\newcommand{\rrem}[1]{$[$\,\ref{rem#1}\,$]$}

\newcommand{\xx}[1]{(\ref{xx#1})}

\newcommand{\ww}[1]{(\ref{ww#1})}

\newcommand{\yy}[1]{(\ref{yy#1})}
\newcommand{\yya}[1]{(\ref{yy1#1})}

\newcommand{\tge}{~Taylor--Goldstein\, equation~}

\newcommand{\xsn} {\smallskip\newline}
\newcommand{\xmn} {\medskip\newline}

\newcommand{\xsl} {\smallskip\linebreak}

\newcommand{\xnl} {\newline}
\newcommand{\xlb} {\linebreak}
\newcommand{\xss} {\smallskip}
\newcommand{\xms} {\medskip}

\newcommand{\play} {\displaystyle}

\newenvironment{eqa}
  {\begin{eqnarray}}  {\end{eqnarray}}
\newenvironment{eqa*}
  {\begin{eqnarray*}} {\end{eqnarray*}}

\newcommand{\I}{\ensuremath{~[\,z_1\,,\,z_2\,]~}}
\newcommand{\OI}{\ensuremath{~(\,z_1\,,\,z_2\,)~}}

\newcommand{\aint}{~\int_{z_1}^{z_2}}  
\newcommand{\bint}{~\int_{z_0}^{z}} 
\newcommand{\cint}{~\int_{z_c}^{z}}

\newcommand{\ub}{$~(\,u\,,\beta\,)~$}
\newcommand{\kc}{$~(\,k\,,\, c\,)~$}
\newcommand{\T} {\textsc{T}}
\newcommand{\K} {\textsc{K}}
\newcommand{\as} {\textsc{s}}
\newcommand{\uc}{\{\,u(z)\,-\,c\,\}}
\newcommand{\us}{\{\,u(z)\,-\,u(z_s)\,\}}

\newcommand{\uh} {\textsc{U}_h}
\newcommand{\uhh} {\,\ddup(z)\,-\,2\,h(z)\,\dum(z)\,}

\newcommand{\du}{u^{\,\prime}}
\newcommand{\dup}{\ensuremath{u_+^{\,\prime}}}
\newcommand{\dum}{\ensuremath{u_-^{\,\prime}}}
\newcommand{\dupm}{\ensuremath{u_\pm^{\,\prime}}}

\newcommand{\ddu}{u^{\,\prime\prime}}
\newcommand{\ddup}{\ensuremath{u_+^{\,\prime\prime}}}
\newcommand{\ddum}{\ensuremath{u_-^{\,\prime\prime}}}
\newcommand{\ddupm}{\ensuremath{u_\pm^{\,\prime\prime}}}
\newcommand{\ddump}{\ensuremath{u_\mp^{\,\prime\prime}}}

\newcommand{\qqdh} {Drazin \& Howard 1966}

\newcommand{\qqdr} {Drazin \& Reid 1981}

\newcommand{\qqfr} {Friedlander 2001}

\newcommand{\qqfh} {Friedlander \& Howard 1998}

\newcommand{\qqsg} {Goldstein 1931}

\newcommand{\qqho} {Howard 1961}

\newcommand{\qqli} {Lin 1955}

\newcommand{\qqgit} {Taylor 1931}

\usepackage[
top=1.0in, bottom=0.8in, 
left=0.875 in, right=0.875 in, 
includefoot]{geometry}
\usepackage{amsmath}
\usepackage{amssymb}
\usepackage{mathrsfs}

\newtheorem{theorem}{Theorem}[section]

\newtheorem{cor}[theorem]{Corollary}

\newtheorem{lemma}[theorem]{Lemma}

\newtheorem{prop}[theorem]{Proposition}

\newtheorem{defn}[theorem]{Definition}

\newtheorem{rem}[theorem]{\textsl{Remark}}

\makeatletter 
\@addtoreset{equation}{section}

\makeatother 

\makeatletter 
\@addtoreset{subsection}{section}

\makeatother 

\begin{document}
{\large{
\title
{Geometry~ of~ Taylor$\,-\,$Goldstein~~equation 
and~stability}
\author{Aravind Banerjee\\
3A/145~ Azadnagar\,, 
Kanpur--208002\,, India\,.\\
aravindban@gmail.com }
\maketitle 
\centerline{\bf{Abstract}}\xms
\tge~(\,TGE\,) governs the stability of 
a shear-flow of an inviscid fluid of 
variable density\,. Using a 
canonical class of its 
transformations it is investigated 
from a geometrical point of view\,. 

Rayleigh's\, point of inflection 
criterion and 
\,Fj\o rtoft's\, condition 
of instability of a homogenous shear-flow 
have been generalized here so that 
only the profile 
carrying the point of inflection 
is modified 
by the variation of ~density\,. 
This fulfils a ~persistent 
expectation in the literature\,. 

A pair of bounds exists such that in 
any unstable flow 
the flow-curvature 
(\,a function of flow-layers\,) 
exceeds the upper bound 
at some flow-layer and falls below 
the lower bound at a higher layer\,. 
This is the 
main result proved here\,. 

Bounds are obtained on the growth rate and 
the wave numbers of unstable modes\,, 
in fulfillment of 
longstanding predictions of ~Howard\,. 
A result of ~Drazin and Howard~ on the 
boundedness of the wave numbers 
is generalized to ~TGE\,. 

The results above hold if the local 
Richardson number does not exceed $~1\,/\,4\,,$ 
otherwise a weakening of the conditions 
~necessary~ for instability\, is seen\,. 

Conditions for the 
propagation of neutrally stable waves 
and bounds on the phase speeds of 
destabilizing waves are obtained\,. 
It is also shown that the 
set of complex wave velocities of normal 
modes of an arbitrary flow is bounded\,. 

Fundamental solutions~ of ~TGE~ are obtained ~and 
their smoothness is examined. 
Finally sufficient conditions for instability
are suggested.
%
%
\section{introduction}

The object of our study here is 
Taylor-Goldstein~ equation (\,TGE\,). It 
was discovered  by  
~G.I.Taylor~  and ~ 
S.Goldstein\,, independently~ and 
simultaneously\,, 
in connection with their studies on 
`\,parallel shear-flow' of a fluid of 
variable density 
(\,see \qqgit ; \qqsg\,)\,. 
It governs the stability 
of the flow\,.~TGE~ is important because of 
its applications to ~oceanography and 
meteorology~ and has received considerable 
attention in the literature 
on hydrodynamic stability 
(\,see~\qqdh; \qqdr\,)\,. 
It describes waves in the ocean and 
clouds in the sky\,. Here we 
explore the fascinating mathematical 
structure\,, that this equation is 
naturally associated with\,. 

In the limit of vanishing 
density variation 
\,TGE\, collapses into the classical 
equation of Rayleigh~
(\,1880\,) which has a well developed~ 
theory 
(~see \qqli\,) characterized 
by its elegance\,, close agreement with
experiments and geometrical flavour\,. 
The fundamental result of this theory 
is the celebrated point of inflection
criterion of Rayleigh for the instability of a 
parallel shear-flow\,.

Theory of shear-flows has many beautiful 
results but much remains 
to be understood yet\,. 
The literature on the subject 
~(\,see~\qqdh ; \qqdr\,)~ is full 
of expectations that under suitable 
conditions the characteristic 
features of ~Rayleigh-theory~ 
will carry over to ~~TGE\,.~There also are 
laments to the effect that the point of 
inflection loses its significance  
in the context of ~TGE\,. In spite of 
persistent and valuable 
attempts starting with 
~Synge~(\,1933\,)\,, Yih~(\,1957\,)\,, 
Drazin~(\,1958\,) 
up to S\,.\,Friedlander~(\,2001\,)\, 
(\,see~\qqdr~;~\qqfr\,)\,, 
these expectations
have found limited fulfillment\,. 
There also are standing questions\,. 
The predictions of ~Howard~(\,1961\,)~ 
are yet to be settled 
and it is not known if the well known 
condition ~(\,Richardson number is 
less than ~1/4~ somewhere in the flow\,)~ 
of \;Howard (\,1961\,) ~and~ Miles (\,1961\,)
is sufficient to ensure 
instability\, (\,see \qqfr\,)\,. Further 
there is no simple 
method for solving \;TGE\,. 
Thus there is a need for a 
detailed investigation\,. 

To understand ~TGE\,, 
we construct a large class of its 
transformations which transform it 
into a canonical form 
resembling ~Rayleigh's equation and 
develop a method that yields 
information on stability via each 
transformation of the class\,. 
The entire 
information is then focused upon the 
question of stability and made 
independent of transformations and 
coherent\,. This results in a 
description of the 
inner core of ~TGE\,. 
A central role in this description is 
seen to be played by a 
characteristic function of the flow\,, 
we call the flow-curvature\,. 

This analysis establishes 
the point of inflection criterion 
for TGE, 
generalizes 
results from ~Rayleigh-theory\,, 
shows that the condition of ~Howard and 
Miles~ does not ensure instability\,, 
proves the predictions of ~Howard~(\,1961\,)\,, 
relates instability to the crossing of a pair of 
bounds in a definite order by the flow-curvature\,,
examines the propagation of neutral and 
unstable waves 
and yields ~fundamental 
solutions of ~TGE\,. 
\begin{eqa}
&\mbox{\tge~~is~~~~~~}
&w^{\prime\prime}(z)~+~A(z)~w(z)~=~0 
~~~~~~~~~~~~~~
\label {xx1} \\ 
&\mbox{with~~boundary~~ conditions~~~~~~~~~~}
& w(z_1) ~=~ 0  ~=~  w(z_2)\,. ~~~~~~~~~~~~         
\label  {xx2}
\end{eqa}
Here   $~~z_1\,\leq z \,\leq z_2\,,~$ 
$~~w(z)~~$  is a complex-valued 
~$~C^{\,2}-$\,function on  ~\I  ,~ the 
prime ~$``~~^\prime~~"$~ denotes ~
differentiation ~$~d/dz~$~ and 

\[
A(z)~=~\,-k^2~-~\frac{u^{\prime\prime}(z)}
{\;~u(z)-c~\,}~+~
\frac{g\,\beta(z)}{~\left\{\,u(z)-c\,\right\}^2~}
\]
where $~~g\,>\,0~~$ is a ~constant\,,  
$~~\beta(z)~=~-\,{\rho^{\,\prime}(z)}/
{\rho_o}~\ge~0$ ~(\,unless stated 
otherwise\,)\,, ~the flow velocity 
$~~u(z)~~$ and the fluid density 
$~~\rho(z)~~$  are sufficiently
smooth ~real--valued functions\,, 
the average density
$~~\rho_o~~$  is a positive constant and ~ 
$~~c\,=\,c_r \,+\, i\,c_i~~$  ~is
a complex constant called the
complex wave velocity\,.~ 
$~~k\,>\,0~~$ is the wave number\,, 
$~~c_i~~$ the phase speed ~and 
$~~k\,c_i~~$ the growth rate of a 
harmonic perturbation wave 
$~~(\,k\,,\,c\,)\,.$ 

The stability of\, 
`\,parallel shear-flows\,'\, is an important
problem in classical theory of hydrodynamics\,. 
Many eminent investigators have contributed
to its understanding (\,see~\qqfh\,)\,. 
TGE~ governs the stability of 
a \;2-dimensional ~parallel shear-flow~
of an inhomogeneous fluid\,. 

We imagine an inviscid and incompressible 
fluid of variable density
flowing in the $~~(\,x\,,\,z\,)$-plane\,, 
between the lines
$~~z=z_1~~$ and $~~z=z_2~$ . 
Here the $~~z-$axis~
is vertical\,, directed upwards and the 
$~~x-$axis~  is horizontal 
$~~\hat{\imath}~~$   
being the unit vector along it\,.  

Suppose at the point 
$~~(\,x\,,\,z\,)~~$  the 
velocity of the fluid particle
is  $~~u(z)\,\hat{\imath}~$ 
and the density is  $~\rho(z)\,.$ 
Thus a layer of the fluid characterized 
by a fixed value of $~z~$ has a constant 
velocity and density\,. It experiences 
a force due to buoyancy\,, proportional to the 
density gradient~ $~\beta(z)~$,~exerted on it 
by the neighbouring layers\,. 
These incompressible and parallel 
layers of the fluid slip smoothly~ on 
each other and generate a shearing 
movement in the fluid\,. 
This time--independent\, 
flow is called a ~parallel shear-flow~ 
and denoted by \ub.

The (\,linear\,) instability of a  
\,shear-flow\, is perceived of in the theory 
of stability\,, 
as the existence 
of a harmonic perturbation wave\,,  
which vanishes at the boundaries\,, 
whose amplitude grows
exponentially with time and whose 
propagation is permitted by 
the linearized form of the
equation of motion of the flow---
a nonlinear PDE 
due to E\"{u}ler\,.

Only stable parallel shear-flows 
are seen to persist in the laboratory 
or in nature\,. Governed by the 
~equation of motion\,, 
the ~unstable flows~ even if they be 
~parallel~ to start with\,, soon ~become 
~time--dependent~ and ~nonparallel\,. 
This is because small and time--dependent 
natural perturbations grow with time 
in these flows\,. 
Thus stability is directly 
related to the question~: 
which parallel flows persist~?

Here we adapt from ~classical hydrodynamics 
the well known criterion 
for the stability of this flow 
(\,see ~\qqdr\,)\,, 
as a definition as follows : 

\textsl{
A harmonic perturbation wave ~
$~(\,k\,,\,c\,)~$~
is called a normal mode of
the flow \ub if 
equations (1) and (2) have a 
nontrivial smooth solution ~$ w(z) $\,. 
A normal mode is called unstable if 
$~~c_i\,>\,0~~$ and neutral if 
$~~c_i\,=\,0\,.$ 
A flow is
called unstable if it has at least
one unstable normal mode and 
stable otherwise.  }

The most ~fundamental result~ on ~TGE~ was 
conjectured by ~Taylor in ~1931~ and 
~proved ~by 
~Howard (\,1961\.) ~and~ Miles 
(\,1961\,) . 
Howard transformed ~TGE~ to prove that 
for any unstable flow ~\ub, 
the ~flow ~discriminant 
$~~\Delta(z)~$ ~is positive 
for some $~~z~~$ in \OI, where 
$\Delta(z)\,=\,
[~u^{\,{\prime}\, 2}(z)~-~
4\,g\,\beta(z)~]~$. 
Equivalently if Richardson number
$~~[~g\,\beta(z)\,/\,
u^{\,{\prime}\, 2}(z)~]
~\ge~1\,/\,4~~$ 
everywhere\,, then the flow is stable\,. 

Synge (\,1933\,)~ proved that 
$~~u_{\,\min}\,<\,c_r\,<\,u_{\,\max}~~$  
for any unstable mode ~\kc~ of a flow ~\ub. 
Howard (\,1961\,)~ went deeper and 
elegantly proved that 
in this case the wave 
velocity $~~c~~$ must lie in the upper 
half plane\,, inside the semicircle 
whose diameter is the real interval 
$~~[\,u_{\,\min}\,,\,u_{\,\max}\,]\,.~$ 
Kochar \& Jain (\,1979\,)~ took a step 
further and replaced the semicircle 
by a semi-ellipse inside it with the 
same base\,. 
Howard also proved that the 
growth rate $~~k\,c_i~~$ 
of unstable modes is bounded above by 
$~~\sqrt{\,\Delta_{\,\max}\,}~/\,2~\,,$
generalizing 
a result due to Hoiland~(\,1953\,) 
(\,see \qqdh\,)\,. 
Further he predicted that $~~k\,c_i\,
\rightarrow\,0~~$ as $~~k\,\rightarrow\,
\infty~~$ 
(\,this is proved easily for homogeneous flows\,)~ 
and also that under suitable 
conditions the wave numbers $~k~$ 
of unstable modes of a flow 
are bounded\,. 

The special case of ~TGE~ when $~~u\,
\equiv\,0~~$ and $~~\beta~~$ takes arbitrary 
real values was known to ~Rayleigh\,. 
In this case instability is expected 
due to gravitational overturning if a 
heavier layer of the fluid 
lies above a lighter layer\,.  
Rayleigh provided the mathematical 
support for this observation and 
proved that the static flow ~~$
(\,0\,,\,\beta\,)~~$ is unstable ~if and 
only if $~~\beta(z)\,<\,0~~$ 
for some $~~z~$. 

We now discuss the basic results on 
~homogeneous flows 
$~~(\,u\,,\,0\,)\,.$ 
When $~~\beta\,\equiv\,0~,~$ \xx1 is 
called ~Rayleigh's equation\,. Rayleigh 
(\,1880\,) proved the fundamental result 
that in any unstable flow\,, $~~\ddu(z)~~$ 
must assume both ~positive ~and~ 
negative~ values~ so that the flow-velocity 
$~~u(z)~~$ must have a point of 
inflection at some point 
$~~z_s~~$ in ~\OI. 
Fj\o rtoft (\,1950\,) added that 
in this case 
$~~\ddu(z)\,\us~<~0~~$ for some $~~z\,.$

Friedrichs (\,1942\,) 
(\,see \qqdh\,)~ using ~Sturm-
Liouville theory proved the existence 
of a smooth\,, neutrally stable mode 
$~~(~k_s\;,\,u(z_s)~)~~$ 
if ~for some $~~z_s~~$ in ~\I,~ 
$~~\ddu(z_s)\,=\,0\,,~~\K(z)\,=\,-\,\ddu(z)\,
/\,\us~~$ is integrable and $~~\K(z)\,>\,
\pi^{\,2}\,/\,(\,z_2\,-\,z_1\,)^{\,2}~~$ 
whenever $~~u(z)\,\ne\,u(z_s)\,.~$ Here\xsn
$~~~~~~~~~
k_s^2~=~-\,\play\min_\phi\,\llm~\aint\,
(~\phi^{\,\prime\, 2}
\,-\,\textsc{K}\,\phi^2~)\,dz~~{\Big /}~
\aint\,\phi^2\,dz~\rrm\,,$ \xsn
the minimum being taken over all functions 
$~~\phi~~$ such that $~~\phi~~$ and 
$~~\phi^{\,\prime}~~$ are square 
integrable and $~~\phi(z_1)\,=\,0\,
=\,\phi(z_2)\,.~$ 

Drazin \& Howard (\,1966\,)~ proved that 
if $~~\K(z)~~$ is integrable and 
$~~\K(z)\,\ge\,0~$ 
whenever $~~u(z)\,\ne\,u(z_s)\,,$ 
then $~~k^2\,\le\,k_s^2~~$
for any unstable mode ~\kc. 
Their argument also shows that 
$~~\K(z)\,>\,
[\,\pi^{\,2}\,/\,(\,z_2\,-\,z_1\,)^{\,2}
\,+\,k^2\,]~$ somewhere\,. 
The last statement follows also from 
independent arguments given by 
M.B.Banerjee \textit{et\,.\,al\,.} 
(\,2000\,)\,.
Thus if $~0\,\le\,\K(z)\,\le\,
\pi^{\,2}\,/\,(\,z_2\,-\,z_1\,)^{\,2}~$ 
whenever $~u(z)\,\ne\,u(z_s)\,,$ 
then the flow is stable\,. 

Heisenberg (\,1924\,) and Tollmien (\,1929\,)~ 
(\,see \qqdr\,)~ 
obtained neutral wave solutions 
$~(\,c_i\,=\,0\,)~$ of ~Rayleigh's~ equation~ 
when $~~u(z)~$ is analytic in 
some neighbourhood of the point 
$~~z\,=\,z_c~~$ in the complex 
$~~z-$plane~ and $~~\du(z_c)\,\ne\,0~~$ 
where $~~u(z_c)\,=\,c\,.~$ 
Their results agree\,. 
They show that one of the solutions is 
analytic near $~~z_c~~$ and vanishes at 
it~ while the other has a 
logarithmic branch point there\,. 

The results proved here will be 
described now\,.
In ~\S\,2~ and ~\S\,3~ we mainly 
examine the stability of flows for which 
$~~\du(z)\,,~\Delta(z)~~$ and 
$~~\beta(z)~~$ are nonnegative functions 
while theorems \tth7~ and ~\tth b~ 
deal with other classes of flows\,. 

Theorem \tth1~ is the generalization of the 
criteria of \,Rayleigh\, and 
\,Fj\o rtoft\,. 
The only difference from 
the homogeneous case as above is that 
the function $~~u(z)~~$ is replaced by 
$~~u_+(z)\,,~$ where~ 
$~~2\;u_\pm^{\,\prime}(z)~=~
[~\,u^{\,\prime}(z)\,\pm\,
\sqrt{\,\Delta(z)\,}~]\,.~$ 
It shows that the condition 
~`$~\Delta(z)\,>\,0~~$ 
for some $~~z~$'~ of ~Howard and Miles 
is not sufficient for instability\,.
It is also shown that 
for unstable modes the 
growth rate 
$~~k\,c_i~\rightarrow~0~~$ as 
$~~k~\rightarrow~\infty\,,~$ and ~that~ if 
$~~\beta(z)\,>\,0~~$ everywhere ~then 
$~~k~~$ is bounded above\,. These prove 
the predictions of ~Howard~(\,1961\,)\,.

Section (3) relates the flow-curvature 
$~~\textrm{T}(z)=
\ddup(z)\,/\,2\,u_-^{\prime}(z)~~$ 
to stability\,.
Theorem \tth2~ gives a pair of bounds 
$~~k\, \coth\,
[\,k\,(\,z-z_i\,)\,]\,,~~i=1,2,~~$  
which $~~\textrm{T}(z)~~$ must cross 
if ~\kc~ is an unstable mode\,. 
In particular $~~\textrm{T}(z)~~$ 
must cross the bounds $~~1\,/\,
(\,z\,-\,z_i\,)~~$ if the flow 
is unstable\,. 
It is shown in ~corollary~\ccor1\; 
that if $~~\textrm{T}(z)~~$ 
is bounded above or bounded below then 
$~~k~~$ is bounded\,. 
Corollary~\ccor2\; shows that for a fixed 
upper or lower bound on ~\T\,, flows with 
sufficiently small depth $~~(\,z_2\,-\,z_1\,)~~$ 
are stable\,.

Theorem \tth3 ~sets an order in which 
the bounds must be crossed by $~~\T(z)~~$ 
in an unstable flow ~\ub~ when $~~\ddu(z)~~$ 
~and~ $~~\beta(z)~~$ have no zeros in common\,. 
If ~\kc~ is an unstable mode then for some 
$~~t_1~<~t_2\,,~$ $~\T(t_i)~=~
k\, \coth\,
[\,k\,(\,z-z_i\,)\,]~~$ for each $~~i\,.$ 
Theorem \tth4~ includes the case when 
$~\ddu(z)~~$ 
and $~~\beta(z)~~$ have common zeros\,. 

Corollary \ccor3~ shows that if for some 
$~~z_s~~$ in ~\I, \T~ is bounded above when 
$~~z\,\le\,z_s~~$ and is bounded below when 
$~~z\,\ge\,z_s\,,~$ then $~~k~~$ is bounded 
for unstable modes ~\kc. Theorem \tth5~ 
leads to the same conclusion when 
~\T~ is bounded below in 
$~~z\,\le\,z_s\,,~$ bounded above in 
$~~z\,\ge\,z_s\,,~$ $~~\ddup(z)~$ 
and $~\beta(z)~$~ have ~no~ zeros in 
common and $~~\du(z_s)\,\ne\,0\,.~$ 
Theorem \tth6~ refines the result when 
$~~\ddup(z)~$ and 
$~\beta(z)~$~ have common~zeros\,.

In ~\S\,\,3.4\, stability of flows 
is examined when $~~\du(z)~~$ and $~~
\Delta(z)~~$ change sign in ~\I in a 
prescribed way (\,see condition B\,)\,. 
Theorem \tth7~ shows that weaker but nontrivial 
necessary conditions of instability 
continue to persist showing thereby 
that some of these flows are stable\,. 
For flows not satisfying condition (B)\,, 
no conditions seem to be necessary 
for instability\,, 
suggesting that all these flows are 
unstable\,.

In ~\S\,3.5\, 
a sufficient condition 
for the propagation of neutral 
waves $\,(~c_i\,=\,0~)\,$ is obtained\,. 
Theorems \tth8 ~and~ \tth9~ give necessary 
conditions for existence of marginally 
stable modes ~$(\,
u_{\,\min}\,\le\,c\,\le\,u_{\,\max}\,)~$ 
~and~ internal gravity waves 
$~(~c\,<\,u_{\,\min}~$ ~or~ $~
c\,>\,u_{\,\max}~)~$ 
respectively\,. It is seen that the 
possibility of propagation of  marginally 
stable neutral waves of arbitrarily large 
wave numbers remains open even if the 
flow-curvature be bounded but the 
set of wave 
numbers of the internal gravity waves is 
bounded if either of the functions 
$~~\ddupm(z)\,/\,2\,u^{\,\prime}_\mp(z)~~$ 
is ~bounded\,.

Upper bounds are obtained 
in ~\S\,3.6~ 
in two different 
situations\,,~on the phase speed 
$~~c_i\,,$ for unstable flows not satisfying  
a \;Rayleigh--Fj\o rtoft\; type condition\,. 
Theorem \tth a~ shows that for a flow 
satisfying condition~ (A) either $~~u_-(z)~~$ 
has a point of inflection at $~~z_s~~$ with 
$~\ddum(z)\,\{\,u(z)-u(z_s)\,\}\,>\,0~~$ 
for some $~~z~~$ or 
$~~c_i\rightarrow 0~~$ as $~~\nu_{\,\max}
\rightarrow 0~~$ where $~\nu^{2}(z)\,=\,
[\,1\,-\,4\,g\,\beta(z)\,
/\,u^{\,\prime\,2}(z)\,]\,.$ 
Theorem \tth b~ shows that for a flow 
with $~\beta(z)\,<\,0~~$ somewhere\,, 
either $~~u_+(z)~~$ 
has a point of inflection at $~~z_s~~$ with 
$~\ddup(z)\,\{\,u(z)-u(z_s)\,\}\,<\,0~~$ 
for some $~~z~~$ or 
$~~c_i\rightarrow 0~~$ as $~~\nu_{\,\max}
\rightarrow 1\,.~$\xss

Propositions \ppn a~ and ~\ppn b\, 
together show that the 
set of complex wave velocities of normal 
modes of an arbitrary flow is bounded\,. 

In ~\S\,4~ ~TGE~ is converted into
a quadratic 
recursion relationship on a sequence of 
smooth functions on ~\I~ and solved\,. 
For a class of flows
Theorem \tth f ~gives a pair of 
linearly independent~ smooth~ solutions~ 
of ~TGE~ in some neighbourhood of the 
critical layer $~~z\,=\,z_c~~$ where 
$~~u(z_c)\,=\,c_r\,,$ when 
$~~c_i\,>\,0~~$ is sufficiently small\,. 
Further if $~~\beta\,\equiv\,0~~$ corollary 
\ccor f~ gets back as a limiting case\,,
the smooth solution 
of ~Rayleigh's equation~ that vanishes 
at $~~z\,=\,z_c~~$ when $~~c_i\,=\,0\,,~$ 
in agreement with results of ~Heisenberg 
and Tollmien (\,see \qqdr\,)\,. 

In ~\S\,4.1~ 
sufficient conditions are suggested 
for the existence of 
an unstable mode solution to the 
~Taylor-Goldstein boundary 
value problem\,. 
\section{The basic method}                    
In this section stability of monotonic 
flows with nonnegative discriminant 
will be discussed\,. 
The basic method is developed in 
~\S\,2\,.\,1\,. It is 
then used to obtain
generalizations of the instability 
criteria of ~Rayleigh ~and~ Fj\o rtoft~ 
in ~\S\,2\,.\,2\,. Sharper use of 
this method will be made in ~\S\,3\, 
to obtain deeper results\,.
\subsection{Canonical 
transformations of ~TGE}
The information on stability obtained 
from transformations of 
~TGE~ is stated in ~lemma \llem1\,. 
It is then restricted to a canonical class 
of transformations to obtain 
~propositions \ppn1 ~and~ \ppn2\,. These 
have been developed in sufficient 
generality in view of their many
applications later in the text\,.
\begin{lemma}     \label                    {lem1}
Let \ub be 
an unstable flow\,.\,
Suppose \,$~f~$\, is a 
complex--valued ~$~C^1-$function\,.
Let $~~f^{\,\prime\prime~}~$ be 
integrable and ~$~f(z)\, \neq\, 0~~$ 
for  every $~~z~~$ in \I~ then 
\[ \aint f^2 \,|\,F^{\,\prime}\,|^{\,2}\,dz~=~
\aint \big( ~f\,f^{\,\prime\,\prime}~+~
A\,f^2 ~\big)\,|\,F\,|^{\,2}\,dz\,. \]
Further ~if 
$~~Im\big\{f^2(z)\big\}\,<\,0~~$ 
for  every ~z~ in \I 
,~ where $~Im~$ denotes 
the ~imaginary ~part~ then 
\begin{eqa*}
Im\,\big[~f(z)\,f^{\prime\prime}(z)
\,+\,A(z)\,f^2(z)~\big]~<~0 &&
~~\mbox{for ~some ~z~ in ~\OI.} 
\end{eqa*} 
\end {lemma}
\textit{\textbf{Proof\,:}}~~
Substituting ~$F(z)~=~
{w(z)}/{f(z)} $~ 
in ~\xx1 gives
\begin{eqa}                        \label   {yy1}
\big[~f^2(z)\,F^{\,\prime}(z)~
\big]^{\,\prime} \,+\, 
\big[~f(z)\,f^{\,\prime\,\prime}(z) \,+\, 
A(z)\,f^2(z)~\big]
F(z)~=~ 0~~~~~~~~~
\end{eqa}
valid at all points $~~z~~$ where 
$~~f^{\,\prime\,\prime}(z)~~$ exists\,.
On multiplying this with $~~ F^*(z)~~$
(~the complex conjugate of ~$F(z)$~) and
integrating one gets
\begin{eqa*}
\aint f^2|\,F^{\,\prime}\,|^{\,2}\,dz\,=
\aint\big(\,ff^{\,\prime\,\prime}\,+
\,Af^2\,\big) |\,F\,|^{\,2}\,dz \,+
\aint\big(\,f^2 F^{\,\prime} 
F^*\,\big)^\prime \,dz\,. 
\end{eqa*}
Now $~~f^2 F^{\,\prime} F^*~=~[\,w^{\,\prime} f 
~-~wf^{\,\prime}\,]~(wf^{-1})^*~~$~ is continuous
and its derivative is integrable in ~\I.~ Further 
$~~f^2 F^{\,\prime} F^*~~$ vanishes
at the boundary points $~~z_1~~$ 
~and~ $~~z_2~$~ by
\xx2\,. Thus the last integral in
the equation above vanishes proving 
the first statement\,. 
The lemma now follows from the imaginary 
part of this equation.

Let the flow-discriminant ~$~\Delta(z)~=~
[~u^{\,{\prime}\, 2}(z)~-~
4\,g\,\beta(z)~]~$.~
In this and the next section we shall
mainly focus upon the class of flows~  
\ub~~satisfying :\xmn
\noindent{\textbf{Condition $(A)$~:}}~~
$~(a)~~~\du(z)\,,\,\beta(z)
\,\ge\,0~~$ for every 
$~~z~~$ in ~\I.\xsn
$~(b)~~~\Delta(z)~~$ 
has a smooth extensions to 
some neighbourhood of ~\I~in 
$~~\mathbb{R}~~$ satisfying 
$~~\Delta(z)\,\ge\,0~~$ everywhere\,.
\begin{rem}
\label {rem1}
\textsl{It is easily seen that if ~$~(~ 
u\,,~\beta\,,~k\,,~c_r\,,~c_i\;;~w~)$~ is a 
solution of the equations ~\xx1~ \& 
~\xx2~ then so is ~
$(~-\,u\,,~\beta\,,~k\,,~-c_r\,,~
c_i\;;~w^*~)$.~Thus either both the flows ~
\ub~and $~(\,-\,u\,,\beta\,)$~are stable or  
both are unstable.~Therefore any result 
which does not change on $~~u~~$ being 
replaced by $~~-\,u~~$ if proved under 
~condition $(A)$\,,~ 
would remain valid~ if the hypothesis~: 
$~~\du(z)~\ge~0~~$ in ~(a)~ above is 
replaced by 
$~~\du(z)~~$~ is 
monotonic in \I.~All the results~ proved 
~under ~condition $(A)$~ 
here~ are ~of ~this ~type\,.}~
\end{rem}

Suppose the unstable flow \ub satisfies 
condition $(A)$\,.~Let
\[u_\pm(z)~=~u(z_1)~+~
\frac{1}{2}\int_{z_1}^z 
\big[~u^\prime(z)\,\pm\; 
\sqrt{\,\Delta(z)\,}~\big] \,dz\,. \]
Part ~(b)~ of condition $~(A)~$ 
ensures that $~~
\sqrt{\,\Delta(z)\,}~~$ is smooth so 
that ~$~u_\pm(z)~~$ are smooth 
~functions ~and ~
$~~2\,u_\pm^{\,\prime}(z)~=~
[~u^\prime(z)\,\pm\,
\sqrt{\,\Delta(z)\,}~]~\geq~0~~$ 
for every $~~ z~~$ in~\I\,. 
~Further 
$~~$if ~$~\beta~\equiv~0~$~ 
then ~$~u_+(z)\,=\,u(z)$\,.

Let ~\kc~ be an unstable mode 
of the flow ~\ub\,. 
The function 
$~~\left\{ u(z)-c~ \right\}~~$ 
then takes its
values in the ~3rd ~and~ 4th~ quadrants.
\begin{eqa}
\mbox{Let}~~~~~~~~~ 
u(z)-\,c &= & |u(z)-\,c\,|~\exp
\left\{\, i\,\theta (z) \;\right\} \label {yy2}\\
\mbox{where}~~~~~~~ -\pi ~ 
< & \theta(z) & <~ 0~.~~~~~~~~~~~~  \label {yy3} 
\end{eqa}
$\theta(z)$~ is then a well defined
smooth function of ~$z$~and a smooth
branch of ~$\log \llm u(z)-c\rrm~ $ 
on \I~ is defined by 
\[\log\llm u(z)-c\rrm
~=~\log|u(z)-c\,|~ +~ i\;\theta(z)\,. \]
On differentiating this equation and taking 
the imaginary part\,, one gets
\begin{eqa}
\theta^{\,\prime} (z)&~=~&
\frac{c_i u^\prime(z)}
{|u(z)-c\,|^2}\label                          {yy4}\\  
&~\geq~& 0~~\mbox
{for~ every ~ z~in ~\I.} \nonumber  
\end{eqa}
$~~~$ Let ~$h(z)$~ be a real--valued\,,~ 
piecewise\,$-C^1$~ function on \I.~
Let ~$~z_o~~$ be ~in ~\I~ and ~let 
\begin{eqa}
f(z)~=~ \exp\llb\frac{i\,\theta(z_o)}{2}  ~+~ 
\int_{z_o}^z\llm\frac{u_-^{\,\prime}(z)}
{~\big\{\,u(z)-c\,\big\}~}~+~ 
h(z)\rrm \,dz~\rrb~.~~~~~~ 
\label                                      {yy5}
\end{eqa}
Clearly ~$f(z)$~ is a 
~$C^1-$function~ and $~~f^
{\,\prime\prime}~~$ is integrable on \I. 
Also ~$f(z)~\neq~0~$ for~ any 
$~~z~~$ in \I.
\begin{eqa}
\mbox{We~~ write}~~~~~~~~~~~
f(z)&=&|\,f(z)\,|~
\exp\,\llb~ i\;\phi(z)/\,2~\rrb~~~~~~~~
\label                                         {yy6}\\
\mbox{where}~~~~~~~~~~~~~~~~~
\phi(z)&=&\theta(z_o)~+~
c_i \int_{z_o}^z\frac{2\,u_-^{\,\prime}(z)}
{~\,|\,u(z)-c\;|^2~\,} \,dz ~~~~~~~~~~~~ 
\label                                       {yy7}\\
\mbox{so~that}~~~~~~~~~~~~~~
\phi^{\,\prime}(z) 
&=&~\frac{\,2\,c_i\,u_-^\prime(z)\,}
{\;~|\,u(z)-c\,|^2~\;}
~~\geq~~0~.~~~~~~~~~\nonumber
\end{eqa} 
It follows now from ~(\ref {yy4})~
and~\yy7~ respectively that 
\begin{eqa}
0~\geq~
\phi^\prime(z)~-~\theta^{\,\prime}(z)
&=&\displaystyle\frac{\,-c_i~\sqrt{\,
\Delta(z)\,}~\,}{\;~|\,u(z)-c\,|^2~}
~~~~\mbox{and}~~~~ 
\phi(z_o)~=~\theta(z_o) 
~~~~~~~~ \label                                {yy8}\\
\mbox{so~that}~~~~~
|\phi(z)-\phi(z_o)|&\leq &
|\,\theta(z)-\theta(z_o)\,|~
~~~~\mbox{for ~every
~z~ in \I.}~~~~ \nonumber
\end{eqa}
Thus ~$\phi(z)$~lies between ~$\theta(z_o)$~
and~~$\theta(z)$\,.~Equations 
~(\ref{yy3}) and (\ref{yy6})~ 
now give in turn 
\begin{eqa}
-\pi~<&\phi(z)&<~0 ~~~~~~~~~~~~~~~~\xss
\label                                       {yy9}\\
\mbox{and}~~~~~~~~~~~~~~
Im\,\llm f^2(z) \rrm 
&=&|f^2(z)|~\sin\llm \phi(z) \rrm~  
\nonumber  \\
&<&0~~~\mbox{for~every~z in \I.}~~~~~~~~~~~  
\label                                       {yy10}
\end {eqa}
It is easily calculated from 
~(\ref{yy5}) that
\begin{eqa}
f(z)f^{\,\prime\prime}(z)+A(z)f^2(z)\,=\, 
\llb~(\,h^\prime+h^2-k^2\,) ~-~
\frac{~~\uh(z)~}
{\,\{\,u(z)-c\,\}\,}~~\rrb f^2(z)
~~~~~~~ \label                                {yy11}
\end{eqa} 
where $~~\textsc{U}_h(z)\,=\,
\big[\,\ddup(z)\,-\,2\,h(z)\,\dum(z)\,
\big]\,.$ 
Equation \yy1 ~then yields 
\begin{eqa*}
~~~\big[\,f^{\,2}(z)\,F^{\,\prime}(z)\,
\big]^\prime + 
\llm(\,h^\prime+h^2-k^2\,)
-\frac{~\textsc{U}_h(z)~}
{~\big\{\,u(z)-c\,\big\}\,}
\rrm f^{\,2}(z)\,F(z)\,=\,0\,. 
\end{eqa*}
On taking the imaginary part of 
\yya1 and using 
\yy2 and \yy6 one obtains
\[ Im\llb~ ff^{\prime\prime}+
Af^2 ~\rrb ~
= \bigg[~(\,h^\prime+h^2-k^2\,)~~
\sin\,\phi\;-\;\frac{~\uh\;
\sin\,(\phi-\theta)~}{|\,u-c\,|}~\bigg] 
\left|f^2\right|\,. \]
In view of ~\yya0 ~and~ lemma \llem 1~ 
it is proved that :
\begin{prop}\label                          {ppn1}
Let \ub be an unstable 
flow satisfying the condition $(A)$\,. 
Let $~h~$ be a real--valued~
,~piecewise\,$-C^1$ 
~function on ~\I~ and let 
~$z_1 \leq z_o \leq z_2 ~$ ~then 
~for ~some $~~z~~$ in ~$(z_1,z_2)$.
\[\frac{~~\uh(z)~~
\sin\,\{\,\phi(z)-\theta(z)\,\}~~} 
{~|\,u(z)\,-\,c\,|~~
\sin\,\phi(z)}~<~
\big[~h^\prime(z)\,+
\,h^2(z)\,-\,k^2~\big]~.~~\]
Further~ suppose $~~
\big[~h^\prime(z)\,+
\,h^2(z)\,-\,k^2~\big]~\leq~0~$ 
everywhere~ 
then the inequality above 
holds at some 
$~z~$ which also satisfies~  
$~\phi(z)\,\neq\,\theta(z)~$ ~and~ 
$~u(z)\,\neq\,u(z_o)~$.
\end{prop} 
\begin{rem}   \label{remz} 
\textsl{
~~If $~~f(z)~~$ and $~~\phi_+(z)~~$ are 
defined by replacing $~~\dum(z)~~$ by 
$~~\dup(z)~~$ in ~\yy5 ~and~ 
\yy7~ respectively then an argument 
similar to the one above yields~:} \xss

\textsl{
If $~-\,\pi\,<\,\phi_+(z)\,<\,0~~$ then 
~proposition \ppn1~ continues to hold 
provided $~~\phi(z)~~$ is ~replaced~ by 
$~~\phi_+(z)~~$ and $~~\uh(z)~~$ 
by $~~\big[\,\ddum(z)\,-\,2\,h(z)\,\dup(z)\,
\big]\,.$ }
\end{rem}

It  will be seen in 
~\S\,3.6\, that violation  of 
the bounds on $~~\phi_+(z)~~$ 
as above implies an 
upper bound on $~~c_i\,.$ 
\begin{defn}       \label                    {def1}
\textsl{
We say that two 
numbers $~~a\,,\,b\, \in \mathbb{R}$\,,
~~are ~sign--equivalent and write  ~$~a\,
\simeq\, b~$~ if 
$~~a\,=\,\lambda~b~~$  for some ~~$\lambda\, 
>\,0~$~~i.e. ~either 
both~~~$a~~and~~b$~~~ are positive 
or both are equal to 
zero or both are negative.}
\end{defn}
Equations \yy3 and \yy9 
show that~$~~ 
-\pi\,<\,\phi-\theta\,<\,\pi$
~~~so~that
\begin{eqa*}
~\sin\big\{\,\phi(z)-\theta(z)\,\big\}~\simeq
~\big\{\,\phi(z)-\theta(z)\,\big\}
~~~~\mbox{for~every~~z~~in~~\I.} 
\end {eqa*}
Further ~\yy8 ~implies that
\begin{eqa*}
~~~\big\{~\phi(z)-\theta(z)~\big\}~\simeq~
-\big\{\,u(z)-u(z_o)\,\big\}~~~~~
\mbox{if}~~~\phi(z)~\neq~\theta(z)~.~~~~~~~~~   
\end{eqa*}
Thus ~if~~
$\sin\big\{\,\phi(z)-\theta(z)\,\big\}~
\neq~0~$~ then
\begin{eqa}
\sin\big\{\,\phi(z)-\theta(z)\,\big\}~
\simeq~-\big\{\,u(z)-u(z_o)\,\big\}~.
~~~~~~~~~~~~~~~~~~~~~~~~~~~~~~
\label                    {yy12} 
\end{eqa}
We now prove :
\begin{prop}             \label              {ppn2} 
~Suppose the flow ~\ub is unstable 
and condition $(A)$\xsl
holds~ then 
$~~\Delta(z)~>~0~~$ for some $~z\,.$ 
Let ~h~ be a ~real-valued\,, 
piecewise\,$-C^1$\xsl 
function~ satisfying ~ 
$\llb~ h^\prime(z)\,+\,
h^2(z)\,-\,k^2~\rrb~\leq~0~~$ for ~every 
$~z~$~in~\I\xsl
and ~let 
$~\uh(z)~=~\big[~\uhh~\big]~~$ then~\xmn
$(a)~~\uh(z)~$ \mbox{takes both ~negative 
and positive~ values~~in~~\I~~so~~that~}\xsn
$~~~~~~~\uh(z_s)~=~o~$~~for~~some 
\mbox{$~z_s~$~in~ \OI}. \xmn 
$(b)~~\,\uh(z)  ~
\llm\, u(z)~-~u(z_s)\,\rrm~ 
<~0~~~for~~some~~z~~in\OI. ~ $
\end{prop}
\textit{\textbf{Proof\,:}}~~
Let $~z_o~$ be any point in \I.
Proposition \ppn1~ shows 
that $~\sin\,(\,\phi-\theta\,)~\neq~0~~$ for 
some $~~z~~$ so that 
$~~\Delta(z)\;>\;0~$~ 
somewhere in ~\I~by ~\yy8\,. 
Further ~\yy9 ~and~  \yya2~ now give
\begin{eqa*}
\uh(z)~\llb~ u(z)-u(z_o)~\rrb
~<~0~~~\mbox{for~~some~~z~~in~}\OI. 
\end{eqa*}
Taking $~~z_o\,=\,z_1~~$ 
(\,resp. ~$~z_o=z_2~~)$ 
here\,,~ we see that ~~$\uh(z)~~$
takes both negative ~(\,resp. positive\,)~ 
values because $~~u(z)~~$ is monotonic.~
This proves part ~(a)\,.~Now ~(b)~ follows on 
substituting $~~z_o=\,z_s~~$ .
\subsection{Generalization of ~
Rayleigh\,,~and~Fj\o rtoft~criteria}
Parts ~(1)~ and ~(2)~ of 
theorem \tth1~ generalize the well known 
instability criteria of~ 
~Rayleigh ~and~ Fj\o rtoft~ while 
parts ~(3)~ and ~(4)~ prove the 
predictions of ~Howard ~(\,1961\,)\,. 
Boundedness of $~~k^2~~$ will be proved 
under weaker conditions in ~\S\,3\,. 
Corollary \ccor a~ shows that the condition 
of ~Howard and Miles~ is not 
sufficient for instability\,.\xss 

Taking ~$h\equiv 0$~ in ~proposition 
\ppn2~ gives the 
~parts ~(1)~ and ~(2)~ below\,. 
\begin{theorem} \label                       {th1}
Let~ the~ unstable~ flow \ub 
satisfy~ condition $(A)$ and \xsl
let \kc be ~one ~of ~its ~unstable~ 
modes~ then \xmn
$~~\,(1)~~\ddup(z)~$ takes~ both~ negative~ and~ 
positive~ values~ so~ that \smallskip\newline
$~~~~~~~~~\ddup(z_s)~=~0~
~for~some~z_s~in~\OI\,, $ 
\[\,~(2)~~\ddup(z)\,\llb~ u(z)-u(z_s)~ 
\rrb<0~~~for~~some~~z~~in~\OI~~\,,~
~~~~~~~~~~~~~~~~~~~~~~~~~\]
$~~(3)~~~k^2\,c_i~<~\play\frac{1}{2}\;
|\;\ddup\;|_{\,\max}~~$~so ~that~~
the~ growth~ rate ~$~k\,c_i~
\rightarrow~0~~~
as~~~k  \rightarrow~\infty~~$,\xmn
~$~~~(4)~~~If ~~\beta(z)\,>\,0~~ 
everywhere~\,in\, \I, 
~\;then\,,~~k^2\,<\,
[~\T^{\,\prime}(z)\,+\,\T^{\,2}(z)~]\xsn
~~~~~~\,~~~for~~some~~z~~in~~\OI,~~where~~ 
\T(z)\,=\,\ddup(z)\,/\,2\,\dum(z)\,.$
\end{theorem}
\textit{\textbf{Proof\,:}}~~
Only the parts ~(3) 
and (4)~ remain to be proved\,. 
It is an easy consequence ~of ~Howard's~ 
semicircle--theorem~ that 
$~~u_{\,\min}~<\, c_r~< 
u_{\,\max}~\,,$ 
(\,a result due to Synge\,)~ 
where~ $~~u_{\,\min}~~$ 
and $~~u_{\,\max}~~$ denote
respectively the minimum and the maximum of 
$~~u(z)~~$ for $~~z~~$ in~\I.~
Let ~~$z_o$~~ be a 
point satisfying $~~u(z_o)\,=\,c_r\,,$ ~then 
~\yy2 shows that 
\[\exp\,\{\,i\,\theta(z_o)\,\}~=~-i\,.\]
Thus ~$~\theta(z_o)~=~-\,{\pi}/\,2~~$ 
~and~ $~\phi(z)~$~ lies~between~~$-\,
\pi/\,2~~$ and 
$~~\theta(z)$~~by 
~\yy8\,. 
Let~ $~~\sin\,x\,=\,|\,\sin\,\theta\,|~~$ 
~and~ $~~\sin\,y\,=\,|\,\sin\,\phi\,|~~$ 
~where $~~0\,\le\,x\,
\le\,y\,\le\,\pi\,/\,2\,.~~$ 
Clearly then $~~\sin\,(y-x)\,=\,
|\,\sin\,\{\phi-\theta\}\,|~~$ 
because both $~~\theta(z)~$ 
and $~\phi(z)~$ 
are either in $~(\,-\pi\,,\,-\pi\,/\,2\,]~$
or in $~[\,-\pi\,/2\,,\,0\,)\,.$~Now \xmn
$~~~~\play\max_x\,\big[\,
\sin\,(x)\;\sin\,(y-x)~\big]\,=\,
\sin^2\,(y\,/\,2)\,=\,\frac{1}{2}\,\sin\,y\;
\tan\,(y\,/\,2)\,
\le\,\frac{1}{2}\,\sin\,y\,,~$ 
\[\mbox{~~~\,so that}~~~~~~~~~
2\,|\sin\,\theta\,|~
|\sin\,(\phi -\theta)\,|~<~|\sin\,\phi\,|~~~~
\mbox{~~for~~ every~~z~~in~~\I.}~\]
Proposition \ppn1~ now shows that 
\[~ 2\,k^2 \,|\;u(z)-c\;|~ 
|\sin\,\theta\,(z)\,|~<~
|\;\ddup(z)\;|~~~~~
\mbox{for~~some~~z~~and~~so} \]
\[2\,k^2\,c_i~<~|\;\ddup\;|_{\,\max}~\equiv~
\max_{z\in \I}{|\;\ddup(z)\;|\,.}~~~~~~~~~\]
This proves part (3)\,. 
Part (4)~ 
follows from ~proposition 
\ppn1\,, on substituting $~~T(z)~~$ 
for $~~h(z)~~$ because $~~
\textsc{U}_T\,\equiv\,0\,.$ 
This completes the proof\,. 

\begin{rem}
\label {rem3}
{\textsl{~~It is easily seen from the theorem 
that the criteria of Rayleigh and 
Fj\o rtoft remain valid\,, as stated by them\,,
when the flow-discriminant $~~\Delta(z)$
is a constant or the Richardson number
~$~R(z)\,=\,g\,\beta \,/\, u^{\prime\, 2}~$~ 
is a constant.}}
\end{rem}

The theorem shows that the flow $~~
(\,2\,g\,z^2\,,~g\,z^2\,)~~$ is stable 
though $~~\Delta(z)~$ 
is positive everywhere\,.
\begin{cor}
\label {cora}
The condition $~~\Delta(z)\,>\,0~~$ 
for some $~~z~~$ in ~\I~ is\xsn 
not ~sufficient ~to ~ensure ~instability\,.
\end{cor}
$\mbox{Further we have} ~~~2\,\ddup(z)\,=\,
\ddu(z)\,+\,\play
\frac{~\du(z)~\ddu(z)~
-2\,g\,\beta^{\,\prime}(z)~}
{~\sqrt{\,\Delta(z)\,}~}\,.~ 
\mbox{This gives}~: $
\begin{cor}      \label   {corb}
~~~Suppose $~\du(z)\,,~\beta(z)\,,~$and
$~\Delta(z)~>~0~$ for every $~z~$ in 
\I, then 
\[ \du(z)~\ddu(z)~\simeq~\beta^{\,\prime}(z)
~~~~~~~~~~~~\mbox{for~~some}~~z~~
\mbox{in}~~\I. \]
\end{cor}
\section{Geometry of \tge}                           
Valuable information on the stability 
of flows satisfying ~condition ~(A)~ 
is coded in the ~propositions 
\ppn1 ~and~ \ppn2~ of the previous section 
via the arbitrary nature of the point 
$~~z_o~~$ and the function $~~h(z)\,.$ 
This information is decoded in this 
section to bring out into the open 
the intimate relationship between 
bounds on the flow 
curvature and stability\,.

Theorems ~\tth2~, ~\tth3 ~and~ \tth4~ show that 
$~~\T(z)~~$ must cross 
the bounds $~~k\, \coth\,
[\,k\,(\,z-z_i\,)\,]\,,~~i=1,2,~$ 
in a definite order 
if ~\kc is to be an unstable mode\,. 
On the other hand ~corollary \ccor3~ and 
theorems \tth5 ~and~ \tth6~ give different 
kinds of bounds on $~~\T(z)~~$ each of 
which ensures the boundedness of the wave 
numbers of the unstable modes\,.

In ~\S\,3.4\, the stability of 
nonmonotonic flows with indefinite 
discriminant is examined\,. Theorem 
\tth7~ establishes that the class of 
flows satisfying ~condition~(\,B\,)~ 
has stable flows in it\,. 

In ~\S\,3.5\, necessary 
conditions are obtained for the propagation 
of neutral waves\,. Theorems \tth a ~and~ 
\tth b~ in ~\S\,3.6\, 
give upper bounds on $~~c_i~~$ 
as an alternatives to 
point of inflection type conditions\,. 
\subsection{Bounds ~on ~the ~flow-curvature~
\boldmath
{$~~\ddup(z)\,/\,2\,u_-^{\prime}(z)$}}
\begin{defn}     \label                {def2}
~~~Let ~$\mathscr T~=~
\{~ z \in \I ~\,|~
\,\ddup(z)\,\neq\, 0\,,~
or~\beta(z)\,\neq 0\,~\}$\,,
and let 
the flow--curvature 
$~~\T:\mathscr T \rightarrow\, 
[\;-\infty,\infty~]$~~~be~ 
defined~ by \xsn
$~~~~~~~~~~~~~~~~~~~~~~~~~~\T\,(z)~
=~[~\ddup(z)\;/\;2\,\dum(z)~]~$. 
\end{defn}

Clearly $~~\T $~~is a continuous function 
in view of ~~$\dum(z)~\geq~0\,.~$ 
It will now be proved that 
in any unstable flow $~~\T(z)~~$ must 
cross a pair of bounds\,.

Let ~$h(z)~=~k\,\coth\big\{
k\,(z-z_1+\epsilon)\big\}\,, ~where 
~\epsilon > 0~$ is sufficiently small\,, ~
then $~~h~~$ is a $C^\infty-$
function~  on~\I~ and~
\[h^\prime(z)+h^2(z)-k^2~=~0~~~
\mbox{for~every~$~~z~~$~in~\I.} \]
Applying now~  
proposition \ppn2 ~part(a) ~one gets
\begin{eqa*}
\T(z)-k\,\coth\{k\,(z-z_1+\epsilon\} 
~>~ o~~~~\mbox{for~some~~z~~in~}\mathscr T\,.
\end{eqa*}
In the same way taking ~$~h(z)~=~k\,\coth
\{k\,(z-z_2-\epsilon\}~~$ gives 
\begin{eqa*}
\T(z)-~k\,\coth\{k\,(z-z_2-\epsilon\} 
~<~ o~~~~\mbox{for~some~~z~~in~}\mathscr T\,.
\end{eqa*}
Now in the limit as ~$\epsilon~
\rightarrow~0~ $~ one 
obtains the following theorem.
\begin{theorem}   \label                      {th2}
Let the flow \ub satisfy 
the condition $(A)$ and let 
$~(\,k,c\,)$\xsn
be an ~unstable~ mode\,,~then 
\[\displaystyle{\limsup_{t\rightarrow z}}
~~\T\;(t)~
\geq ~k~\coth~\{~k~(z-z_1)~\}
~~~~~for~~some~~z~~in~~
\overline{\mathscr T}~~~and \]
\[\displaystyle{\liminf_{t\rightarrow z}}
~~\T\;(t)~
\leq ~k~\coth~\{~k~(z-z_2)~\} 
~~~~~for~~some~~z~~in~~
\overline{\mathscr T}\,.~~~~~ \]
\end{theorem}
\begin{rem}
\label {rem4}
\textsl{
~Suppose $~~\ddup(z)~~$
~and~ $~~\beta(z)~~$ have no 
common zeros in ~\I\,, then
$~~\T~~$ is continuous on 
~\I~ ~and the left hand 
sides of ~the ~inequalities~ 
in the ~theorem~ reduce to 
$~~\T\,(z)~$.~}
\end{rem}

Theorem \tth2 ~has 
some~interesting ~consequences.~
Using ~the ~inequality 
~$~\coth\,(t)\,>\,1~~$ 
if $~~t\,>\,0~$~ one obtains :
\begin{cor}   \label                          {cor1}
$~$ Suppose the hypothesis 
of theorem \tth2~ 
holds\,.~Let $~~\T\,(z)$ \xnl
be~ bounded~ above~ 
or~ bounded~ below~ ,~then 
\begin{eqa*}
~~k&<&\min{~\llb~ \displaystyle   
\sup_{z\,\in\,\mathscr T} 
{\llm ~\T\,(z)~ \rrm}
~\,,~~-\inf_{z\,\in\,\mathscr T}
{\llm ~\T\,(z)~\rrm}~\rrb}~~
<~~\infty~
\end{eqa*}
so that the set of wave numbers 
of ~unstable modes~ of ~\ub~ is\, 
bounded\,.
\end{cor}
In ~\S\,3.2 ~and~ 
~\S\,3.3~ boundedness of $~~k~~$ will 
be proved under weaker conditions\,.
Using $~~~k~\coth\,(k\,t)~>~1\,/\,t~~$ 
~for~ $~~t~>~0~~$ ~one obtains :
\begin{cor}  \label                           {cor2}
$~~$Suppose the hypothesis of 
~corollary \ccor1~
holds\,,~then
\begin{eqa*}
(a)~~~~\frac{1}{z-z_1} 
&<& ~~\T\,(z)~~~for~~
some~~z~~in~\OI~~~~and~~~ \\
(b)~~~~\frac{1}{z_2-z} 
&<& -~\T\,(z)~~for~~
some~~z~~in~\OI~~so~~that\\
(c)~~~\frac{1}{z_2-z_1}&<&~~
\min{~\llb~ \displaystyle   
\sup_{z\,\in\,\mathscr T} 
{\llm ~\T\,(z)~ \rrm}
\,,~-\inf_{z\,\in\,\mathscr T}
{\llm ~\T\,(z)~\rrm}~\rrb}~~<~~\infty\,. 
\end{eqa*}
\end{cor}

This shows that given a fixed upper or lower 
bound on $~~\T(z)~~$ flows with sufficiently 
small depth $~~(\,z_2\,-\,z_1\,)~~$ 
are stable\,.~Thus we see that instability 
needs sufficient room to manifest itself. 
\subsection{The main instability criterion}

We are now in a position to prove the 
main result on instability\,. 
\begin{theorem}    \label                    {th3}
Let \ub be an unstable flow 
satisfying the condition $(A)$\,.\linebreak
~Suppose 
$~\ddup(z) ~$
and $~\beta(z) ~$,~have no 
common zeros in \I ~then 
for some $~t_1,~t_2~$ in \I,~ $~t_1 < t_2\,,$
\[\T\,(t_1)~=~ k~\coth 
\big[~k~(\,t_1\,-\,z_1\,)~\big]~>~\frac
{1}{~\,t_1\,-\,z_1\,}~~ \]
\[\mbox{and}~~~~~~~~~~~~ 
\T\,(t_2)~=~ k~\coth 
\big[~k~(\,t_2\,-\,z_2\,)~\big]~<~\frac
{1}{~\,t_2\,-\,z_2\,}~~.~~~~~~~~~~~~~~~~ \]
\end{theorem}
\textit{\textbf{Proof\,:}}~~
Remark~\rrem4 ~
shows that $~~\T~$,~
in this case is a continuous 
function into $~[~-\infty,\infty~]~$. 
\[Let~~~~~~t_1 = \displaystyle
{\min_{z_1\le\, z\le\, z_2}
~{\big\{~ z~|~~{\T(z)}}~
\geq~ k~\coth~[~k~(z-z_1)~]~\big\}}\]
\[and~~~~~~t_2 = \displaystyle
\max_{z_1\le\, z\le\,z_2}  
~{\big\{~ z~|~~{\T(z)}~
\leq~ k~\coth~[~k~(z-z_2)~]~\big\}\,.}\]
Theorem \tth2~ shows~ that~ 
$~~t_1 ~~and~~~ t_2~~$~ 
are~ well~ defined\,,~and~ clearly 
\[~t_1 ~\neq~ t_2~~~~\mbox{and}~~~~ 
\T(t_i)~=~k~\coth~[~k~(t_i-z_i)~]~~~
\mbox{for} ~~i=1,2\,.~\]
We prove now that $~t_1~<~t_2~$.~ 
If possible suppose ~$t_2<t_1$\,.~
Let ~$z_s$~ be any 
point in ~$(\,t_2\,,\,t_1\,)$~\,,~
and ~$~\T_s~=~
\T(z_s)~$.~
Let ~$\epsilon~>~0~$ ~ 
be a small number and
let$~~~h:\I~\rightarrow 
\mathbb R~~$ 
be defined by
\[ h(z)~=~\left\{
\begin{array}{l l}
k\,\coth\;[~k\,(\,z-z_1+\epsilon\,)~]
~~~&\mbox{if}~~~~~~
z~\in~[~z_1\,,~z_s-\epsilon~]~~; \\
l_1(z)~~~~&\mbox{if}~~~~~~
z~\in~[~z_s-\epsilon\,,~z_s~]~~;\\ 
l_2(z)~~~~&\mbox{if}~~~~~~
z~\in~[~z_s\,,~z_s+\epsilon~]~~;~\\ 
k\,\coth\;[~k\,(\,z-z_2-\epsilon\,)~]
~~~&\mbox{if}~~~~~~
z~\in~[~z_s+\epsilon\,,~z_2~]~~; \\
\end{array} \right. \]
where the graphs of $~l_i(z)~$~
for $~~i=1,2~~$ are obtained 
by joining the point \xlb
$~~(\,z_s\,,~\T_s \,) ~~~$
with the points~~~
$\big(~z_s \mp \epsilon\,,~k\,
\coth\;[~k\,(\,z_s-z_i\,)~]~\big)~~$
respectively\,,~ by line segments\,.
When $~\epsilon~$ 
is sufficiently small 
it is easily checked that\,,~
\begin{description}
\item 1. $~h~~$ 
is a piecewise\,$-C^1$~ function\,,~~and~~ 
$~h(z_s)~=~\T(z_s)~$;
\item 2. $~h^\prime +h^2-k^2~=~0~$ if $~z~$ is 
not in $~[~z_s-\epsilon\,,~z_s+\epsilon~]~$;
\item 3. $~k\,\coth\;
[~k\,(\,t_2-z_2\,)~]~\,\leq\,~h(z)~\,
\leq\,~k\,\coth\;[~k\,(\,t_1-z_1\,)~]~$ 
for~ every $~z~$ in ~
$[~z_s-\epsilon\,,~z_s+\epsilon~]~;$
\item 4. $~$for any fixed~ $~m\,<\,0~$,~~ 
$~h^\prime (z)~<~m~$~ for every $~z~$ 
in $~[~z_s-\epsilon\,,~z_s+\epsilon~]~;$
\item 5. $~h^\prime +h^2-k^2~\leq~0~$~ for 
every $~z~$ in \I;
\item 6. $~\T^{\,\prime}(z)
\,-\,h^\prime(z)~>~0$ 
~~for~ every~ $~z~$ in ~
$[~z_s-\epsilon\,,~z_s+\epsilon~]~;$  
\item 7. $~\T(z)~<~h(z)~$ \,if\, 
$~z~\leq~z_s-\epsilon~$~~ and ~ 
$~\T(z)~>~h(z)~$ \,if\, 
$~z~\geq~z_s+\epsilon~$;
\item 8. $~\T(z)= h(z)~~$ only if 
$~~z=z_s~~$ and 
$~\T(z)-h(z)~\simeq~ (z-z_s)~$.  
\end{description}
$\mbox{It follows that}~~~
\big[~\T(z)-h(z)~\big]~\big[~u(z)-
u(z_s)~\big]~\geq~0~~~
\mbox{for~ every}~ z~\mbox{in} \I.$ \xsn
This contradicts ~proposition \ppn2\,.~
Thus ~$t_1<t_2~$,~and the proof is 
complete\,.

With a little more care~ one can include 
the case when $~\ddup(z)~$ and $~\beta(z)~$ 
have common zeros\,.~
We state the result~~
and briefly indicate its proof\,.
\begin{theorem}    \label                     {th4}
Let \ub be an unstable flow 
satisfying condition $(A)\,,$ 
then for 
some ~$~t_1~$ and $~t_2~$~ in~~ 
$\overline{\mathscr T}~$,~
$~t_1~\leq~t_2~$, 
\end{theorem}
$~~~~~~~~~(\textsc A)~~~
\displaystyle\limsup_{z\rightarrow t_1}
~\T(z)~\geq ~ k~\coth 
\big[~k~(t_1-z_1)~\big]~
>~\frac{1}{t_1-z_1}~~~~and~ $ \xnl
$~~~~~~~~~~(\textsc B)~~~~
\displaystyle\liminf_{z\rightarrow t_2}
~\T(z)~\leq ~ k~\coth 
\big[~k~(t_2-z_2)~\big]~
<~\frac{1}{t_2-z_2}~~~~ $ \xmn 
\textit{and ~one ~of ~the ~
following ~conditions ~(a) ~
or ~(b) ~hold\,: }\xmn
(a)~~ $~t_1~<~t_2~$;~\xmn 
(b)~~$~t_1~=~t_2~=~t_0~$
(\,\textit{say}\,)~ 
\textit{and ~one ~of ~the ~
following ~conditions ~hold\,:} \xmn 
$~~~~$(1)~~$~\textsc l_+\,\ge\,
k~\coth 
\big[~k~(t_1-z_1)~\big]~~~$ where ~~
$\textsc l_+~=~\displaystyle
\limsup_{z\rightarrow t_0-0} 
~\T(z)~.$  \xmn
$~~~~$(2)~~$~\textsc r_-\,\le\;
k~\coth 
\big[~k~(t_2-z_2)~\big]~~~$ \,where ~~
$\textsc r_-~=~\displaystyle
\liminf_{z\rightarrow t_0+0} 
~\T(z)~.$  \xmn
$~~~~$(3)~~$~\textsc l_+~>~
\textsc r_-~.~$\xmn
$~~~~$(4)~~$~\textsc l_+~=~
\textsc r_-~=~
\T_0~$(say)~~~~ and~~ 
$\displaystyle 
~~~\liminf_{z\rightarrow t_0}~
\frac{\T(z)-\T_0}{z-t_0}~~
=~~-\infty~.$  \xmn
\textit{\textbf{Proof\,:}}~~
\[Let~~~~~~~~~t_1 =~ \displaystyle{\inf
~{\big\{~ z~\in~\mathscr T~|~~
\displaystyle\limsup_
{t\rightarrow ~z}{\T(t)}}~
\geq~ k~\coth~[~k~(z-z_1)~]~\big\}}~~~~\]
\[and~~~~~~~~~t_2 = \displaystyle{\sup  
~{\big\{~ z~\in~\mathscr T~|~~
\displaystyle\liminf_
{t\rightarrow ~z}{\T(t)}}~
\leq~ k~\coth~[~k~(z-z_2)~]~\big\}}~.~~~\]
It is clear that inequalities ~
(\textsc A) ~and~ (\textsc B)~ 
above hold\,.

Suppose both conditions 
~(a) ~and~ (b)~ do not hold\,.
Clearly ~$~t_2~\leq~t_1~$. 
If $~t_2~=~t_1~$ then~~ 
$~k~\coth\,[\,k~(t_2-z_2)\,]~<~
\textsc l_+~\leq~\textsc r_-~<~
k~\coth\,[\,k~(t_1-z_1)\,]\,$
and ~$~t_1~$~ is not in $~\mathscr T~$. 
Let $~~z_s\,=\,[\,t_1\,+\,t_2\,]\,/\,2~~$
and ~let \xmn
$~~~~~~~\T_s~=~\left\{
\begin{array}{l l l c c}
\T(z_s)~~~&
\mbox{if}~~~&t_2\,<\,t_1
~~~&\mbox{and}~~&z_s~~
\mbox{is~ in }~\mathscr T \\
\lls~\textsc l_+\,+\,\textsc r_-\,
\rrs/\,2
~~~&\mbox{if}~~~&t_2\,=\,t_1
~~~&\mbox{or}~~&z_s~~
\mbox{is~ not~ in }~\mathscr T\,.  
\end{array} \right.$ \xss

Let the function $~h~$ be defined as 
in the proof of theorem \tth3~ above\,.~
It is not difficult to check\,, as in 
the proof of the previous theorem\,, that 
for sufficiently small $~\epsilon~$
\[\big[~\T(z)-h(z)~\big]~\big[~u(z)-
u(z_s)~\big]~\geq~0~~~~~~~~
\mbox{for~ every}~ z~\mbox{in} \I. \]
This contradicts ~proposition \ppn2\,.
Thus condition\,(a)\,
or \,(b)\, above\, must\, hold\,.\xss

For any $~~z_s\;,t_1\;,t_2~~$ in ~\I, 
$~~t_1\,\le\,t_2\,,$~ either $~~t_1~~$ 
is in $~~[\,z_1\,,\,z_s\,]~~$ ~or~ 
$~~t_2~~$ 
is in $~~[\,z_s\,,\,z_2\,]\,.~$ 
From parts ~(A) ~and~ (B)~ of the 
~theorem ~ it follows easily that :
\begin{cor}        \label               {cor3}
~~Let \ub be ~an ~unstable ~flow 
~satisfying ~condition $(A)$\,, 
and let $~~z_s~~$ be a point in 
\I\,. Suppose $~~\T(z)~\le~\T_0~~
in~~z~\le~z_s~$ 
$and~~\T(z)~\ge~-\T_0~~
in~~z~\ge~z_s\,~$ for some 
$~~T_o\,>\,0\,,~$ then 
$~~ k~<~\T_0~~$ 
and ~the ~set ~of  
wave numbers 
~of ~unstable modes~ of ~\ub~ is 
bounded\,.
\end{cor}
\subsection{Bounds on the 
wave numbers of unstable modes .}

Corollary \ccor3~ gives conditions on 
$~~\T(z)~~$ so that the unstable modes 
have bounded wave numbers\,.
In this subsection different sets of 
conditions will be described that 
ensure the same\,. 
The following lemma generalizes a deft 
piece of estimation in 
Drazin \& Howard (\,1966\,)\,.~
Let $~z_s \in$ \I ~~and~ let \xnl$~$\xnl
$u_0~=\left\{\begin {array}{lcl}
2c_r-u(z_s)   ~~~~~&\mbox{if}&~~~~~  
u(z_1)~\leq~ 2\,c_r-u(z_s)~
\leq~ u(z_2)~;\xss \\ 
u(z_1) ~~~~~&\mbox{if}&~~~~~ 
2\,c_r-u(z_s)\, ~\leq~\, u(z_1)~;\xss \\
u(z_2) ~~~~~&\mbox{if}&~~~~~
2\,c_r-u(z_s)\, ~\geq~\, u(z_2)\,.
\end{array} \right.$\xss
\begin{lemma}   \label   {lem6} 
Let $~z_s \in$ \I ~~and~$~u_0~$ be as 
above\,.~Let $~z_o~$ be a point 
such that ~$~u(z_o)~=~u_0\,.~$
Let ~$~\theta(z)~~and~~ \phi(z)~$~
be as in ~proposition \ppn1\,,~
then $~\eta(z)~\leq~1~$
for every $~z~$ in ~\I~ ~where
\[\eta(z)~=~
\frac{~u(z)-u(z_s)~}{~|\,u(z)-c\,|~}\,.~
\frac{~\sin\big\{\phi(z)-\theta(z)\big\}~}
{\sin\phi(z)}~. \]
\end{lemma}
\textit{\textbf {Proof\,:}}~~
let $~z_c~$ be a point 
such that $~u(z_c)=c_r\,.$~
For any $~t~$ ,~we write $~u_t~~and~~
\theta_t~$ ~for~ $~u(z_t)~~
and~~\theta(z_t)\,.$ 
We assume first that ~$~u_s
\leq c_r~$,~then~\xss\newline
$~~~~~~~~~~~~z_s~\leq~ z_c~
\leq~ z_o~~$ ~~~and~~~~
$\theta_s ~\leq~ -{\pi}/{2}~\leq~ 
\theta_0 ~\leq~ 
-\pi -\theta_s\,.$\xss\newline
Case(1)~:~ $~z_1~\leq~ z~\leq~ z_s~$,~
then \smallskip~\newline 
$~~u(z)~\leq~ u_s~~$~~ and 
$~~~~~0~\leq~ \phi(z)-\theta(z)~
~\leq~ \pi+\theta(z) ~\leq~ -\theta_s
~\leq~ \pi~$,
\begin{eqa*}
\mbox{so that}~~~~~~~
c_i~\cot(\phi-\theta)~&\geq &~ 
-c_i~\cot(\theta)~=~u_s-c_r~. \\
\mbox{Thus}~~~~~~~~~~~~~~~~~~~~~~
\eta(z) &=& \frac{(u-u_s)}
{~|\,u-c\,|~ 
\{~\cos\theta-sin\theta~ \cot(
\phi-\theta)~\}}\\
&=&\frac{(u-u_s)}{~(u-c_r)-c_i~\cot(
\phi-\theta)}\\
&\leq& \frac{(u_s-u)}
{~(c_r-u)+(u_s-c_r)}~=~1\,. 
\end{eqa*}
Case(2)~:~ $z_s~\leq z~
\leq z_o~$,~then\newline
$u-u_s~\geq 0~\,,~~0~\leq \phi-\theta~
\leq \pi~$~~
and~~ $\sin\phi ~\leq 0$~~~~so that~~~  
$\eta(z)~\leq 0~.$\medskip \newline
Case (3)~:~ $z_o~\leq z~\leq z_2~$,~then 
\newline\smallskip
$(u-u_s)~\geq~ 0~~~~and~~~~
0~\geq~ \phi-\theta~ 
\geq~ \theta_0~=~-\pi-\theta_s~\geq~-\pi~$
\begin{eqa*}
\mbox{and so}~~~~~~~~~
c_i~\cot(\phi-\theta)&\leq&~c_i~\cot
(-\pi-\theta_s)~~
=~(u_s-c_r)~.~~~~~~\\
\mbox{Thus}~~~~~~~~~~~~~~~~~~~~~\eta(z)&=&
\frac{u-u_s)}{(u-c_r)-c_i~\cot
(\phi-\theta)}\\
&\leq & \frac{(u-u_s)}{(u-c_r)-(u_s-c_r)}
~=~1\,.
\end{eqa*}
We have proved that~~if $~~u_s~\leq~c_r~~$
then $~~\eta(z)~\leq~1 ~$~ for every $~~z\,.$ 
A similar argument applies when 
$~u_s~\geq~c_r~$ ~and completes the 
proof of the lemma\,.
\begin{prop} \label        {ppn3}
~Let \ub be an ~unstable 
flow ~satisfying condition $(A)$\,.\xsn
Let $~h~$~ be a 
~piecewise\,$-C^1$ 
~function~ on~ \I. 
Let $~~z_s~~$ be~ in~ \I \xsn
~and~ $~\uh(z_s)\,=\,0\,.$~ 
Let ~~$~
\textsc{K}_h(z)\,=\,-\,\uh(z)~  / ~
\{\;u(z)\,-\,u(z_s)\;\}~~$ 
whenever \xsn 
~$~~u(z)\;\neq\;u(z_s)~~$ and~ 
suppose $~~\textsc{K}_h(z)\;\ge\;0~~$. 
Let $~~k^2~~$ be~ 
sufficiently~ large \xsn
so that 
$\llb~ h^\prime(z)\,+\,
h^2(z)\,-\,k^2~\rrb~\leq~0 $ 
~~for~every~$~z~$~in~\I\,,~then~ 
\[k^2~<~\displaystyle {\sup_z~
\big\{~h^\prime(z)\,+
\,h^2(z)\,+\,\textsc{K}_h(z)~\,
|\,~{u(z)\,\neq\, u(z_s)}~\big\} }\]
and~ so~ the~ set~ of~ 
wave~ numbers~ of~ the~ 
unstable~ modes~ is~ bounded~ 
if \xsn
$~\textsc{K}_h(z)~$ is~ bounded~ on~ the~ 
set~ $~\{~u(z)\,\neq\, u(z_s)~\}$.
\end{prop}
\textit{\textbf {Proof\,:}}~~
Let $~z_o~$ be as in ~lemma \llem6\,.~
From ~proposition \ppn1~ we have \xsn
$~~~~~~~~~~~~~~~
\textsc{K}_h(z)\,.~\eta(z)~<~h^\prime(z) 
+h^2(z)-k^2 ~~
\mbox{~for~ some }~~z~~ in~ \I. $ \xsn
The proposition now follows from 
~lemma \llem6\,.
\begin{rem}
\label {rem5}
\textsl{~~Suppose 
$~~\du(z_s)\,\neq \,0\,,~$ 
then $~~\textsc{K}(z)~~$ 
has a continuous 
extension to \I,} 
and~ so~ it~ is~ bounded\,. 
\end{rem}

Taking $~~h\,\equiv\,0~~$ in this 
proposition we obtain a generalization 
of a result due to ~Drazin \& Howard~
(\,1966\,)\,.
\begin{cor}   \label                          {cor4}
~~Let ~$~
\textsc{K}_0(z)\,=\,-\,\ddup(z)~  / ~
\{\;u(z)\,-\,u(z_s)\;\}\,\ge\,0\,,~$ 
whenever  
~$~~u(z)\;\neq\;u(z_s)\,,~$ then 
$~~k^2\,<\,\textsc{K}_0(z)~~$ for some $~~z\,.$
\end{cor}

We use this proposition to obtain the 
following intrinsic 
criterion for the boundedness 
of the wave numbers of unstable modes\,.
\begin{theorem}     \label                    {th5}
~~Let \ub be \,an \,unstable 
\,flow \,satisfying \,condition $(A)$\,.\xsn
Suppose ~$~\ddup(z)~$ 
and $~\beta(z)~$~ 
have ~no~ common zeros\,. 
Let ~$z_1 \leq z_s \leq z_2 $ \xss\xlb
and ~let~~$~\du(z_s)\,\ne\,0\,.~$
Let~ $~~\T\,(z)~>~
\T_1~~~~if~~~~ z\,<\,z_s~~$ and\xsl
$ \T(z)~<~\T_2~~~~if~~~ z\,>\,z_s~~$ 
for ~some~ constants ~$~T_1~$ 
and $~T_2~$, then\xsl
the~ set~ of~ wave~ numbers ~of 
the~ unstable~ modes~ is~ bounded\,.
\end{theorem}
\textit{\textbf {Proof\,:}}~~
It ~is ~clear ~from ~the ~hypothesis ~that 
$~~\T(z_s)~~$ is finite\,. 
Let $~~T_s\,=\,\T(z_s)~~$ and let 
$~~\epsilon >0~~$ be a small number\,.~
Let $~h~$ 
be the piecewise linear function on \I\,,~
whose graph is obtained by joining the 
points ~$~(\,z_1\,,\,\T_1\,)\,,~~
(\,z_s\,-\,\epsilon\;,\,\T_1\,)\,,~
~(\,z_s\,,\,\T_s\,)\,,~~
(\,z_s\,+\,\epsilon\;,\,\T_2\,)~~~and ~~
(\,z_2\,,\,\T_2\,)~$.

Clearly $~~h(z_s)\,=\,\T(z_s)\,.$ 
When $~\epsilon~$ is 
sufficiently small ~and~ 
$u(z)\,\ne\,u(z_s)\,,~$ we have
\begin{eqa*}
\frac{\T(z)-h(z)}
{u(z)-u(z_s)}\,
>\,0~~~~~
\mbox{so that}~~~~~
\textsc K(z)\,=\,\frac
{~-\,\uh(z)~}
{~u(z)-u(z_s)~}\,>\,0\,. 
\end{eqa*}
Further by the remark above\,, 
$~~\textsc K(z)~~$ 
is bounded above\,. The theorem now 
follows from ~proposition \ppn3\,.

With some more effort one can 
include the case when $~~\ddup(z)$~~
and~~$\beta(z)~~$ have common zeros\,. 
We state the result\,.
\begin{theorem}     \label                    {th6}
~Let ~\ub be~ an ~unstable 
flow ~satisfying~ condition $(A)$\,.\xsn
Let ~$z_1 \leq z_s \leq z_2 ~$. 
Let~~$~\du(z_s)\,\ne\,0\,.~$
Let~ $~\T\,(z)~$ 
~satisfy  
\end{theorem} 
$\T\,(z)~>~
\T_1~~~~if~~~ z<z_s~~$
~~~~and~~~~~~$ \T(z)~
<~\T_2~~~~
if~~~ z>z_s~~~~$\newline
$~~~~~~$for~~ some~~ constants~~
$\T_1$~~and~~$\T_2$\,.
\smallskip \newline
(b) $\textsc{L}_-\geq~\textsc{R}_+~$,~~~~
where~~~$~\textsc{L}_-~=~\displaystyle
\liminf_{z\rightarrow z_s-0}~
\T(z)~$~~and~~~
$ \textsc{R}_+~=~
\displaystyle\limsup_{z\rightarrow z_s+0}
~\T(z)$ \smallskip\newline
(c) If~~$\textsc{L}_-~=~\textsc{R}_+~=
~\T_s~$(say)~~\,,~ then~~ 
$\displaystyle 
\limsup_{z\rightarrow z_s}~
\frac{\T(z)-\T_s}{z-z_s}~~
\ne~\infty\,.$ \smallskip\newline
\textit{~~~~~~Then~ the~ set~ of~ 
wave~ numbers~of ~the~ 
unstable~ modes~ is~ bounded~}.
\subsection
{Flows with indefinite velocity 
and discriminant\,.}
$~~$
In this subsection the stability 
of a class of flows\,,~in which 
$~\du(z)~$ 
and~ $~\Delta(z)~$ ~change sign in \I ~
in a prescribed way will be discussed\,. 
It is shown that 
nontrivial necessary conditions of 
instability continue to persist\,,~though in 
a weaker form\,,~showing thereby that  
some of  these flows 
are stable\,.~This class of~ 
flows~ is~ described~ by 
the following condition\,.\xmn
\noindent{\textbf{Condition(B):}}
~~A flow \ub satisfies 
this ~condition~ 
if for some $~s_1\,,~s_2~$ 
in \I,~$~s_1\,\le\,s_2~$,\xsn
(1)~~$\beta(z)~\geq~0~$ 
for every $~z~$ in \I.~\xsn
(2)~~$\Delta(z)~~$ has an extension to 
some neighbourhood of ~\I satisfying 
$~~~~~~~\Delta(z)~\leq~0~$ 
if $~z~$ is in 
$~[~s_1\,,\,s_2~]~$~ and 
$~~\Delta(z)\,\ge\,0~~$ otherwise\,.\xsn
(3)~\,$~u(z)~$ is ~monotonic ~in 
~each ~of ~the ~intervals\,, 
$~~[\,z_1\,,\,s_1\,]~$ ~and~ 
$[\,s_2\,,\,z_2\,]~$.\xss

Thus~ 
$~\Delta(s_i)~=~0~$,~ 
for $~i=1,2~$ ; ~  $~\du(z)~$
is unrestricted in  
$~[~s_1,s_2~]\,,~$ 
and ~its~ signs in 
$~~S_1\,=\,[\,z_1\,,\,s_1\,]~$ ~and~ 
$~~S_2\,=\,[\,s_2\,,\,z_2\,]~$
could be different\,. 
For $~~i\,=\,1\,,2~~$ let 

\[ \sigma_i\,=\,\left\{
\begin{array}{r l}
1~~~~&\mbox{if} 
~~~\du(z)\,\ge\,0~~
~~~\mbox{for~ every~} ~z~ \mbox{~in~} 
~S_i~; \\
-1~~~~&\mbox{if} 
~~~\du(z)\,\le\,0~~
~~~\mbox{for~ every~} ~z~ \mbox{~in~} 
~S_i\,.
\end{array} \right. \]
\[Let~~~~~~~
u_\pm(z)\,=~u(z_1)~+~
\frac{1}{2}\int_{z_1}^z 
\big[~u^\prime(z)\,\pm\; 
p\,(z)~\big] \,dz ~\,,
~~~~~~~~~~~~~~~~~~~~~~~~~\]
\[ where~~~~~
p\,(z)\,=\,\left\{
\begin{array}{c l}
\sigma_i\,\sqrt{\,\Delta(z)\,}~~~~
& \mbox{if ~~$\Delta(z)~\ge~0~$}~ 
\mbox{~~and~~} z~~ 
\mbox{is~ in~} ~~S_i~;\\
0~~~~
& \mbox{if ~~$\Delta(z)~\le~0~$}\,.
\end{array} \right. \] 

This extends the definition of 
$~~\ddupm(z)~~$ given earlier 
after ~remark \rrem1\,. 
$u_{\pm}(z)~~$ are $~C^1$--functions 
in ~\I\,,~ 
and have continuous second derivatives 
except possibly at the points 
$~~s_1~~$ and $~~s_2\,.~$ We now prove :
\begin{theorem}                      \label       {th7} 
Let \ub be an unstable 
flow satisfying 
condition (B)\,,\,then\xsn 
$(a)~~(-)^{i}\,
\sigma_i\,\ddup(z)\,<\,0~~$ 
for some $~i\,=\,1~or~2~$ ~and~ some 
$~~z~~$ in $~~S_i~.$ \xmn
$(b)~\;$If~ $~\Delta^\prime(s_1\,-\,0)
~\ne~0~\ne~\Delta^\prime(s_2\,+\,0)~~$ 
then $~~u_+(z)~~$ 
has ~a ~point ~of \xlb
$~~~~~~$ inflection~ 
at~ some~ point $~~z_s~~$ outside 
$~~[\,s_1\,,\,s_2\,]\,,$ satisfying 
$~~\Delta(z_s)\,\le\,0\,.$\xmn
$(c)~$ If 
$~~s_1\,=\,s_2\,,~$ let 
~one  ~of ~the ~
~conditions ~$(1)$ ~
or ~$(2)$ ~hold :\xmn
$~~~~~~~~\,~~(1)~~~\textsc l_+\,<\,
\textsc r_-~~$
where ~~
$\textsc l_+\,=\,\displaystyle
\limsup_{z\rightarrow t_0-0} 
~\T(z)~$ ~~and~~ 
$\textsc r_-\,=\,\displaystyle
\liminf_{z\rightarrow t_0+0} 
~\T(z)\,,$ \xnl
$~~~~~~~~~(2)~~~\textsc l_+\,=\,
\textsc r_-\,=\,
\T_0~$(say)~~~~ and~~ 
$\displaystyle 
~~~\liminf_{z\rightarrow t_0}~
\frac{\T(z)-\T_0}{z-t_0}~
\ne~-\infty\,,$  \xsn 
$~~~~~$ then ~
one  ~of ~the ~
~conditions ~$(3)$ ~
or ~$(4)$ ~is ~satisfied :\xmn
$~~~~~~~~~~(3)~~~
\displaystyle{\limsup_{t\rightarrow z}}
~~\T\,(t)~\geq ~k\,
\coth\,\{\,k\,(\,z\,-\,z_1\,)\}
~~~for~~some~~z~~in~~
[~z_1\,,\,s_1~]\,, $\xmn
$~~~~~~~~~~(4)~~~
\displaystyle{\liminf_{t\rightarrow z}}
~~\T\,(t)~\leq ~k\,
\coth\,\{\,k\,(\,z\,-\,z_2\,)\}
~~~for~~some~~z~~in~~
[~s_2\,,\,z_2~]\,, $\xmn
$~~~~~~$so ~that ~if $~~(-)^{\,i+1}\,\T(z)~~$ 
is bounded above in $~~S_i~~$ for each 
$~~i\,,~$ then\xmn
$~~~~~~~~~~~~~~~
k~<~\max{~\llb~ \displaystyle   
\sup_{z\,\in\,S_1} 
{\llm ~\T\,(z)~ \rrm}
~\,,~~-\inf_{z\,\in\,S_2}
{\llm ~\T\,(z)~\rrm}~\rrb}~~
<~~\infty~.$ 
\end{theorem}
\textit{\textbf {Proof\,:}}~~~~~~
Let $~~z_o~=~
[\,s_1\,+\,s_2\,]\,/\,2\,.$ 
Let $~~\phi(z)~~$ be as in ~
\yy7\,, with $~~u_\pm(z)~~$ as 
~defined~ above\,. Clearly~ 
$~~\phi^{\,\prime}(z)\,=
\,\theta{\,^\prime}(z)~~$~ 
in $~~[\,s_1\,,\,s_2\,]\,.$ 
It follows that 
$~~\phi(z)\,=\,\theta(z)~~$~ 
in $~~[\,s_1\,,\,s_2\,]\,.$ 
Further if $~~z~~$ is ~in $~~S_i~~$ and 
$~~\phi(z)\,\ne\,\theta(z)\,,$ then~ 
$~~[~\phi(z)\,-\,\theta(z)~]~
\simeq~(-)^{i+1}\,\sigma_i~$  
~[~see ~definition~~\ddef1~]\,, 
because $~~\phi^{\,\prime}(z)\,-
\,\theta{\,^\prime}(z)~\simeq~
2\,\dum(z)\,-\,\du(z)~\simeq~-\sigma_i\,.$~

It ~follows ~from 
~proposition \ppn1~ ~and~ 
equation \yy9\,, that ~for ~any 
~piecewise--$C^1~$ function 
$~~h(z)\,,~$ satisfying 
$~~\llb~ h^\prime(z)\,+\,
h^2(z)\,-\,k^2~\rrb~\leq~0~~ $ 
and \mbox{ 
for ~some} $~~i\,=\,1$~or~$2~$
\begin{eqa}
\label {ww1}
~~~~~~(-)^{i}\;
\sigma_i\;\uh(z) ~<~0~~\mbox{~~for~~ some} 
~~z~~ \mbox{in} ~~S_i\,.
\end{eqa} 
Part (a)~ follows ~from ~this ~on ~taking 
$~~h~\equiv~0\,.~$ 

Under the conditions of part (b)\,,  
$~~{~\sigma_1\;\ddup(z)~}
~\rightarrow~-\infty~$ 
as $~~z\rightarrow 
s_1-0\,,~$ ~and 
${~\sigma_2\;\ddup(z)~}
~\rightarrow~\infty~$ 
as $~~z\rightarrow ~s_2+0\,.$ Now 
part (b) ~follows ~from ~~(a)\,. 
To prove part (c)~ we may assume that 
if$~~s_1\,=\,s_2~~$ then \xsn
$k~\coth\,[\,k~(s_2-z_2)\,]~<~
\textsc l_+~\leq~\textsc r_-~<~
k~\coth\,[\,k~(s_1-z_1)\,]\,.$~~Let\xsn
$~~~~~~~\T_0~=~\left\{
\begin{array}{l l l c c}
\T(z_o)~~~&
\mbox{if}~~~&s_1\,<\,s_2
~~~&\mbox{and}~~&z_o~~
\mbox{is~ in }~\mathscr T \\
\llb~\textsc l_+\,+\,\textsc r_-~
\rrb\,/\,2
~~~&\mbox{if}~~~&s_1\,=\,s_2
~~~&\mbox{or}~~&z_o~~
\mbox{is~ not~ in }~\mathscr T\,.  
\end{array} \right.$ \xss

Let $~~\epsilon\,>\,0~~$ be a small 
number\,. Let $~~h(z)~~$ be the function 
defined in the proof of theorem \,\tth3~ 
with $~~z_o~~$ and $~~\T_0~~$ replacing 
$~~z_s~~$ and $~~\T_s~~$ respectively\,. 
If possible~ suppose neither of the 
conditions ~(3)~ and ~(4)~ 
hold\,. It is clear then that for 
sufficiently small $~~\epsilon~,~$ \xsn
$~~~~~~~~~
\T(z)\,-\,h(z)\,<\,0~~$ in $~~S_1~~$ ~and~ 
$~~\T(z)\,-\,h(z)\,>\,0~~$ in $~~S_2\,.$ 
\xmn 
Further $~~\sigma_i\,\du(z)\,\ge\,0~~$ 
for every $~~z~~$ outside $~~[\,s_1\,,\,
s_2\,]\,.$ It follows that 
$~(-)^{i}\;
\sigma_i\;[\,\uhh\,]\;>\;0~~$ 
for each $~~i\,=\,1$ ~and~ 2~ ~and~ every 
$~~z~~$ in $~~S_i\,.~$ 
This ~contradicts~ \ww1~ 
and ~proves part (c)\,.
\subsection{Propagation of neutral modes}

A normal mode \kc of a flow \ub is 
called a neutral mode if $~~c_i\,=\,0\,.$ 
It is called a marginally stable mode if 
$~~c~~$ is ~in 
$~~[\,u_{\,\min} \,,\,u_{\,\max} \,]\,.$ 
These modes are important\,. In the space 
of normal 
modes of flows where a point looks like 
$~~(\,u\,,\,\beta\,,\,k\,,\,c\,)~~$ 
satisfying  ~\xx1 and \xx2\,, 
it is the marginally stable modes that 
separate stability from instability\,.

A ~neutral ~mode ~with $~~c~~$ ~outside 
$~~[\,u_{\,\min} \,,\,u_{\,\max} \,]~~$ 
~is ~called ~an internal~ 
gravity wave\,. Lemma \llem1 
~with $~~f(z)\,=\,\{\,u(z)-c\,\}~~$ 
then gives  
\[\aint (\,u\,-\,c\,)^{\,2}\,
|\,F^{\,\prime}\,|^{\,2}\,
d\,z\,= \aint \llb-\,k^2\,(\,u\,-\,c\,)^{\,2}
\,+\,g\,\beta(z)\,\rrb\,
|\,F(z)\,|^{\,2}\,d\,z~.~\]
This shows as observed in ~
Drazin \& Howard (\,1966\,)~ that 
no internal gravity waves can exist if 
$~~\beta(z)~~$ is negative everywhere\,.
Otherwise using the well known inequality that 
if a $~C^1-$function $~~\psi(z)~~$ satisfies 
$~~\psi(z_1)\,=0\,=\psi(z_2)~~$ then
\begin{eqa}
\aint |\,\psi(z)\,|^{\,2}\,d\,z\,\le\,
\frac{~(\,z_2-z_1\,)^{\,2}}{\pi^{\,2}}
\aint |\,\psi{\,^\prime}(z)\,|^{\,2}\,dz~~,~
~~~~
\label {ww2}
\end{eqa}
one ~obtains~~~~~$|\,u(z)\,-\,c\,|_
{\,\min}^{\,2}~<~
g\,\beta_{\,\max}\,(z_2\,-\,z_1\,)^{\,2}
\,/\,\pi^{\,2}\,.~$ It follows that :
\begin{prop}   \label {ppna}
Let ~\kc~ be ~a ~neutral 
~mode ~of ~a ~flow ~\ub\, and~ 
suppose~ $~~\beta(z)~\ge~0~~$ 
for some $~~z\,,~$ then \xsn
$~~~~u_{\,\min}\,-\,\alpha~<~c~<~
u_{\,\max}\,+\,\alpha~~ $
~~where~~ $~~\alpha \,=\,
\sqrt{\,g\,\beta_{\,\max}}~~
[\,z_2 \,-\,z_1\,]\;
\big{/}\;\pi~.~~$
\end{prop}

Let $~~c\,\in\,\mathbb R\,.~$ Let 
$~~\beta(z_s)\,=\,
\beta^{\,\prime}(z_s)\,=\,0\,=\,\ddu(z_s)~$ 
whenever $~u(z_s)\,=\,c\,.$ 
Let $~~\play\textsc{K}(z)\,=\,
-\,\frac{\,\ddu(z)\,}{~\uc~}\,+\,
\frac{~g\,\beta(z)~}{~\uc^2~}\,,~$ 
be integrable in ~\I. \xnl
This is the case if $~~c~~$ is outside 
$~~[\,u_{\,\min} \,,\,u_{\,\max} \,]~~$ 
otherwise the condition 
$~\du(z_s)\,\ne\,0~~$ whenever 
$~~u(z_s)\,=\,c~~$ 
is sufficient to ensure this\,. 
Suppose $\;\;\textsc{K}(z)\,>\,\pi^2\,/\,
\{\,z_2\,-\,z_1\,\}^2~~$ whenever $~~u(z)\,
\ne\,c\,.~$ 
Equations\; \xx1 and \xx2 \;then constitute 
a regular Sturm-Liouville problem\,, 
and so the~flow~\ub 
admits a neutral mode 
$~~\lls\,k_s\,,\,c\,\rrs,\,$ where \xsn
$~~~~~~~~~
k_s^2~=~-\,\play\min_\phi\,\llm~\aint\,
(~\phi^{\,\prime\, 2}
\,-\,\textsc{K}\,\phi^2~)\,dz~~{\Big /}~
\aint\,\phi^2\,dz~\rrm~>~0~,$ \xsn
the minimum being taken over all functions 
$~~\phi~~$ such that $~~\phi~~$ and 
$~~\phi^{\,\prime}~~$ are square 
integrable and $~~\phi(z_1)\,=\,0\,
=\,\phi(z_2)\,.~$ 
Further if \kc is any neutral 
mode then $~~k^2\,\le\,k_s^2\,.~$ 
We now prove :
\begin{theorem}   
\label {th8}
~Let \,\kc\, be a smooth neutral 
mode for \ub 
satisfying ~condition $(A)$\,. 
For some $~~z_s\,\in\,\I~~$ let 
$~~u\,(z_s)\,=\,c\,,~~\beta(z_s)\,=\,0\,=
\,\ddup(z_s)$ and 
$~~\du(z_s)\,\ne\,0\,,~$ then ~one ~of ~
the ~following conditions ~hold :\xsn
$~~(1)~~
\displaystyle\limsup_{t\rightarrow z}
~\T(t)~\geq ~ k~\coth 
\big[~k~(z-z_1)~\big]~
>~\frac{1}{z-z_1}~\,,~~
for~~ some~~z~<~z_s $\,. \xnl
$~~(2)~~
\displaystyle\liminf_{t\rightarrow z}
~\T(t)~\leq ~ k~\coth 
\big[~k~(z-z_2)~\big]~
<~\frac{1}{z-z_2}~\,,~~
for~~ some~~z~>~z_s\,. $ \xsn
$~~(3)~~\;
Condition~~(b)~~ of~~ ~theorem~ \tth4~~ holds 
~~~with ~~~~t_1\,=\,t_2\,=\,t_0\,=\,z_s\,.~~$

\end{theorem}
\textit{\textbf {Proof\,:}}~~ 
$~\dum(z_s)\,=\,0~~$ because 
$~~\beta(z_s)\,=\,0\,.~$ 
Let $~~h(z)~~$ be piecewise$-C^1$\,.
\[ \mbox{Let}~~~~~~
f(z)~=~ \exp\llb~
\int_{z_s}^z\llm\frac{u_-^{\,\prime}(z)}
{~\big\{\,u(z)-c\,\big\}~}~+~ 
h(z)\rrm \,dz~\rrb~.~~~~~~ \]
$f(z)~~$ is then smooth and 
positive everywhere\,. 
Equation \yy1 ~then ~shows ~that ~for ~
every $~~z~~$ in ~\I,~ 
\[ \big[~f^2(z)\,F^{\,\prime}(z)~
\big]^{\,\prime} \,+\, 
\llb~\{\;h^\prime(z)+h^2(z)-k^2\;\} ~-~
\frac{~\uh(z)~}{~\{\,u(z)-c\,\}~}~\rrb
F(z)~=~ 0~\,. \]
Multiplication ~by $~~F^{\,*}(z)~~$ 
and ~integration 
~as ~in ~lemma \llem1 ~yields 
\begin{eqa*}
\frac{~\uh(z)~}{~\{\,u(z)-u(z_s)\,\}~}~<~
\{\;h^\prime(z)+h^2(z)-k^2\;\}~~\mbox{
for ~some }~~z\,\in\OI\,.
\end{eqa*}
The theorem follows from this by an argument 
similar to that used to prove theorem \tth4\,. 
We ~next ~prove :
\begin{theorem} 
\label {th9}
~~Let ~\ub~ satisfy ~condition $(A)$\,. 
Let ~\kc~ be ~a 
smooth ~neutral ~
mode ~and~ let $~~c~~$ be outside 
$~~[\,u_{\,\min} \,,\,u_{\,\max} \,]\,,$ 
then \xmn
$~~(1)~~\play
\frac{\ddup(z)}{~2\,\dum(z)~}~\ge~
k~\coth\,\big[~k~(z-z_1)~\big]~~~for~~some~~
z~~~~if~~~~c\,>\,u_{\,\max}\,.$\xsn
$~~(2)~~\play
\frac{\ddup(z)}{~2\,\dum(z)~}~\le~
k~\coth\,\big[~k~(z-z_2)~\big]~~~for~~some~~
z~~~~if~~~~c\,<\,u_{\,\min}\,.$\xmn
$~~(3)~~\,Parts~(1)~~and~~(2)~~
hold~~if~~u_+(z)~~
and~~u_-(z)~~~are~~interchanged\,.$ \xmn
$~~(4)~\,~$If ~either ~of 
$~~[~\ddupm(z)\,/\,2\,u^{\,\prime}_\mp(z)~]~~$ 
is ~bounded ~then $~~k~~$ is ~bounded\,.
\end{theorem}
\textit{\textbf {Proof\,:}}~~
$ \mbox{Let}~~~~~~\play
f(z)~=~ \exp\llb~
\int_{z_s}^z\llm\frac{u_\mp^{\,\prime}(z)}
{~\big\{\,u(z)-c\,\big\}~}~+~ 
h(z)\rrm \,dz~\rrb~~$ in turn\,. 
Parts (1) and (2) follow on 
taking the $~~-ve~~$ sign above~ 
by an argument similar to that used 
to prove theorem \tth3~ 
while part (3) follows on 
taking the $~~+ve~~$ sign\,. 
Part ~(\,4\,)~ follows from ~(\,1\,)\,, 
(\,2\,) and (\,3\,)\,.
\subsection{Bounds on the phase speed 
~{\Large{\textsl c}}$_{i}~$ 
of unstable modes\,.}
This section deals with two situations 
where upper bounds on the 
phase speed $~~c_i~~$ 
are obtained as alternatives to 
~Rayleigh--Fj\o rtoft 
type conditions\,. For flows 
satisfying condition ~(A)\,, ~theorem \tth a~ 
is obtained if $~~u_-(z)~~$ is replaced 
by $~~u_+(z)~~$ in ~ \yy7\,. 
Theorem \tth b~ is about flows in which 
$~~\beta(z)~~$ is negative somewhere\,. In 
both cases an upper bound on $~~c_i~~$ 
is obtained if 
~\yy9~ does not hold\,.\xms

Let $~~~~A\,=\,\pi\,+\,\theta(z_1)\,=\,
\tan^{-1}~[~c_i\,/\,(c_r-u_{\,\min})~]~\,,$\xsn
$~~~$ and $~~~\,B\,=\,~~~-\,\theta(z_2)\,=\,
\tan^{-1}~[~c_i\,/\,(u_{\,\max}-c_r)~]~\,.$\xsn
Clearly $~~0~<~A\,,~B~\le~ \pi\,/\,2\,.~$ 
Let $~~~r\,=\,\frac{1}{2}\,
[\,u_{\,\max}\,-\,u_{\,\min}\,]~$ ~then~ 
$~~c_i\,\le\,{r}\,.~$ \xsn
It follows from ~\yy4~ that 
\begin{eqa}
\theta(z_2)\,-\,\theta(z_1)\,=\,
\pi\,-\,A\,-\,B~=~ 
c_i \aint\frac{~\du(z)~d\,z~}
{~~|\,u\,-\,c \,|^{\,2}~}\,.
~~~~~~~~~\label {ww3} \\
\mbox{Further}~~ 
~~~~~~\min\,\{\,A\,,\,B\,\}~\ge~\tan^{-1}~
\lls\,\frac{c_i}{\,2\,r\,}\,\rrs
\ge~\play\frac{~c_i~}{r}\,
\tan^{-1}~(1/2)\,~~~~~~~~\label{ww4} 
\end{eqa}
because~~ 
$~~\tan^{-1}\,x~\ge~(\,x\,/\,x_o\,)~
\tan^{-1}\,x_o~~$ ~~~if~~~ 
$~~0~\le~x~\le~x_o~\le~\pi\,/\,2\,.~$ 
\begin{eqa}
\mbox{and} ~~~~A\,+\,B 
~=~\play\tan^{-1}~\llm\frac{2\,r\,c_i}
{~r^2\,-\,
\llb~c_r\,-\,\frac{1}{2}
(\,u_{\max}\,+\,u_{\min}\,)
~\rrb^2\,-\,c_i^2~}\rrm ~~~\nonumber\\ 
\ge~\tan^{-1}~(\,2\,c_i\,/\,r\,)~\ge~
\play\frac{~c_i~}{r}\,
\tan^{-1}~(2)\,\,.~~~\,
~~~~~~~~~~~~~~~~~~~ \label{ww5}
\end{eqa}
\begin{theorem} \label {tha}
~~Let ~\kc~ be ~an ~unstable ~mode ~for 
~the ~flow  ~\ub \xsn 
satisfying ~condition $\,(A)$\,. Let 
$~~\nu(z)\,=\,\Delta(z)\,/\,\du(z)~~$ 
and $~~\nu_m\,=\,\play\max_z ~\nu(z)\,,$ \xnl
\mbox{then~~one~~of~~the~~following~~
conditions~~(a)~~or~~(b)~~holds~~}\xmn
$(a)~~~~~
\ddum(z)~$ takes~ both~ negative~ and~ 
positive~ values~ so~ that 
$~~\ddum(z_s)~=~0~~$
$~~~~~~~~~for ~~some ~~~z_s~~~in~~\OI, ~~and~~~ 
\ddum(z)\,\llb~ u(z)-u(z_s)~ 
\rrb\,>\,0~~for \xnl
~~~~~~~~~some~~z~~in~\OI\,.~$ \xsn
$(b)~~
\play c_i\,\le\,\frac{\nu_m\,\pi\,r}
{~[~\tan^{-1}(\,1\,/\,2\,)~+~\nu_m\,
\tan^{-1}\,(\,2\,)~]}~~$ 
so ~that $~~c_i\,\rightarrow\,0~~$ 
as $~~r\,\nu_m\rightarrow 0~.$ 
\end{theorem}
\textit{\textbf {Proof\,:}}~~ 
Let ~$~~\phi_+(z)~=~\theta(z_o)~+~\play
c_i \int_{z_o}^z\frac{2\,u_+^{\,\prime}(z)}
{~\,|\,u(z)-c\;|^2~\,} \,dz~~$ and suppose 
the condition\,:~ 
$~~-\,\pi~<~\phi_+(z)~<0~~$ for ~
every $~~z_o\,,z~~$ in ~\I,~ holds. 
In this case it is easily seen that 
$~~\sin\,\{\,\phi_+(z)\,-\,\theta(z)\,\}
\,\simeq\,\{\,u(z)\,-\,u(z_o)\,\}\,.~$ 
Remark \rrem z~ 
then leads to 
the result in ~proposition \ppn1~ 
with $~~u_+(z)~~$ and 
$~~u_-(z)~~$ interchanged and $~~\phi~~$ 
replaced by $~~\phi_+~$\,. The arguments used 
to prove parts ~(1) 
and (2)~ of theorem \tth1~ now prove the part 
~(a)~ above\,. \xss

On ~the ~other ~hand 
~suppose ~for ~some $~~z_o\,,~z~~$ in ~\I,~ 
the condition\,: $~~-\,\pi~<~\phi_+(z)~<0~~$ 
does ~not ~hold\,, ~then 
\begin{eqa*}
c_i\aint~\frac{~\du\,(\,1\,+\,\nu\,)~d\,z~}
{|\,u\,-\,c \,|^{\,2}} &\ge & 
\min\,(\,A\,,\,B\,)\,+\,
\theta(z_2)\,-\,\theta(z_1) \\
\mbox{so~~ that}~~~~~~~~~~~~~~
\nu_m\,(\,\pi\,-\,A\,-\,B\,)&\ge&
\min\,(\,A\,,\,B\,)~~~~~~~~~~~~~~~~~~~
\end{eqa*} 
\mbox{because ~of ~\ww3\,.} 
Part ~(b)~ now ~follows ~
from ~\ww4 ~and~ \ww5\,. This 
completes the proof of the theorem\,.\xss

The flows ~\ub~ with $~~\beta(z)\,<\,0~~$ 
somewhere will be considered now\,. 
In this case $~~\nu_m\,>\,1\,.$ All these 
flows are believed to be unstable\,. 
We now prove that for such a flow the complex 
wave velocities of unstable modes is bounded\,.
\begin{prop}         \label{ppnb}
~Let \kc be an unstable 
mode of a flow ~\ub~ and 
$\beta(z)~\le~0~~$ 
somewhere\,. Let 
$~~r_o^2\,=\,r^2\,+\,
[\,-\,g\,\beta\,]_{\,\max}~(\,z_2\,-\,z_1\,)^2 
\,/\,\pi^2\,,~$ then \xsn
$(1)~~u_{\,\min}~<~c_r~<u_{\,\max}~~$ and \xmn
$(2)~\llb~c_r\,-\,\frac{1}{2}
(\,u_{\max}\,+\,u_{\min}\,)
~\rrb^2\,+\,c_i^2~\le~r_o^2~\,.$ 
\end{prop}
\textit{\textbf {Proof\,:}}~~ 
Taking $~~f(z)\,=\,\uc~~$ in ~lemma \llem1~ 
and proceeding 
as in the proof of Howard's semicircle 
theorem ~(\,see \qqho\,), one obtains  \xsn 
(1)~~~~$\play \aint \{\,u(z)\,-\,c_r\,\}~
(~|\,F^{\,\prime}\,|^{\,2}
+k^2|\,F\,|^{\,2}~)~d\,z~=~0~$~~~~~and \xsn
(2)~~
$~~\llm\llb~c_r\,-\,\frac{1}{2}
(\,u_{\max}\,+\,u_{\min}\,)
~\rrb^2\,+\,c_i^2\,-r^2\rrm \play \aint
(~|\,F^{\,\prime}\,|^{\,2}
+k^2|\,F\,|^{\,2}~)~d\,z~~$\xsn
$~~~~~~~~~~\le~-\play
\aint g\,\beta \,|\,F\,|^2~d\,z~\le~
[\,-\,g\,\beta\,]_{\,\max}\play \aint
|\,F\,|^{\,2}~d\,z~. $ \xsn
The ~proposition ~now ~follows ~from 
~these and ~\ww2\,. 
\begin{rem}
\textsl{
It is clear from ~propositions \ppn a ~and~ 
\ppn b~ together with ~Howard's semicircle theorem 
that for an arbitrary flow ~\ub, the set 
of complex wave velocities of its normal modes 
is bounded\,.}
\end{rem}
\begin{theorem}    \label {thb} 
~~Let ~\kc~ be ~an ~unstable ~mode ~for 
~the ~flow ~\ub. 
Let $~~\du~~$ and $~~\Delta~~$ be ~nonnegative 
~functions and ~let $~~\beta(z)$ be 
~negative ~somewhere\,. 
Let $~~\nu_c\,=\,\nu(z_c)~~$ 
where $~~u(z_c)\,=\,c_r~~$ and ~let 
$a\,=\,\play\max_z~
\{~\nu^{\,\prime}(z)\,/\,\du(z)~\}~~$ 
{then~~one~~of~~the ~~
conditions~~(a)~~or~~(b)~~holds\,.~}\xmn
$~~~~(a)~~Parts ~~(1)~ 
~and~ ~(2)~ ~of~ 
~theorem~ \tth1~ ~hold\,.~~~~~~~~~$\xmn
$~~~~\,(b)~~~~~~\play c_i~\le~
\frac{(\,\nu_m\,-\,1\,)\,\pi\,r_o}
{~~[~\tan^{-1}(\,r_o\,/\,2\,r\,)~
+~(\,\nu_m\,-\,1\,)
\tan^{-1}\,(\,2\,r_o\,/\,r\,)~]~}
~~~~~~and $ \xsn 
$~~~~~~~~~~~~~~\play\nu_c~\ge~1~-~
\frac{2\,(\,\nu_m\,-\,1\,)\,a\,r_o}
{~~[~\tan^{-1}(\,r_o\,/\,2\,r\,)~
+~(\,\nu_m\,-\,1\,)
\tan^{-1}\,(\,2\,r_o\,/\,r\,)~]~}~\,,$\xmn
$~~~~~~~~~$
so ~that $~~c_i\,\rightarrow\,0~~$ 
~as~~ $~~a\,r_o\,(\,\nu_m\,-\,1\,)\,
\rightarrow\,0\,.~$ 

\end{theorem}
\textit{\textbf {Proof\,:}}~~ 
Let $~\phi(z)~$ be ~as ~in 
~\yy7\,. 
Suppose ~the ~condition\,: 
$~-\,\pi~<~\phi(z)~<0~$ for ~
every $~~z_o\,,z~~$ in ~\I,~ holds\,. 
In this case the same argument that is used 
to prove parts ~(1) and (2)~ of theorem \tth1\,, 
proves part ~(a)\,.\xss

On the other hand 
suppose for some $~~z_o~~$ 
and $~~z~~$ in ~\I,~ 
the condition\,: $~~-\,\pi~<~\phi(z)~<0~~$ 
does not hold  
\begin{eqa}
\mbox{then}~~~~~~~~~~~~~~~
c_i\aint~\frac{~\du\,(\,1\,-\,\nu\,)~d\,z~}
{|\,u\,-\,c \,|^{\,2}} &\le& 
-\,\min\,(\,A\,,\,B\,)~~~~~~~~~~~~~~ 
\label{ww6}\\
\mbox{so~ that}~~~~~~~\,
(~\nu_m\,-\,1~)~(~\pi\,-\,A\,-\,B~)&\ge& ~~
\min\,(\,A\,,\,B\,)~.~~~~ \nonumber
\end{eqa}
The definition of the angles ~A ~and~ B~ 
given above\,, gives in this case 
\begin{eqa*}
~~~~~~\min\,\{\,A\,,\,B\,\}~\ge~\tan^{-1}~
\lls\,\frac{c_i}{\,2\,r\,}\,\rrs
\ge~\play\frac{~c_i~}{r_o}\,
\tan^{-1}~(\,r_o\,/\,2\,r\,)\, \\
\mbox{~~~~~and}~~~~~~~~~~~~~A\,+\,B 
~\ge~\tan^{-1}~(\,2\,c_i\,/\,r\,)~\ge~
\play\frac{~c_i~}{r_o}\,
\tan^{-1}~(\,2\,r_o\,/\,r\,)\,\,,~\,
\end{eqa*}
$~\mbox{so~ that}~~\play
(\,\nu_m\,-\,1\,)\,\pi~\ge~\frac{c_i}{r_o}~
[~\tan^{-1}(\,r_o\,/\,2\,r\,)~
+~(\,\nu_m\,-\,1\,)
\tan^{-1}\,(\,2\,r_o\,/\,r\,)~]~.$

The first inequality in part ~(b)~ follows 
now from this equation\,. To obtain the 
second inequality\,, sharper estimation of 
the integral in ~\ww6~ is needed\,. 
It may be assumed that 
$~~\nu_c\,<\,1\,.~$ 
Let $~~s\,=\,(\,1\,-\,\nu_c\,)\,/\,a~$~ then\xsn 
$~~~~~~~~~~~~~
\play\nu(z)\,-1~<~\nu_c\,-\,1\,+\,a\,s\,=\,0~~~~
~~\mbox{if}~~~~|\,u(z)\,-\,c\,|\,<\,s~~$ \xsn
Let $~~~E\,=\,\play
\frac{1}{2}\,\pi\,-\,\tan^{-1}\,
(\,\frac{c_i}{s}\,)\,.~$~ 
Equation ~\ww6~ then gives 
\begin{eqa*}
(\,\nu_m\,-\,1\,)\,
(\,\pi\,-\,A\,-\,B\,-\,2\,E\,)~\ge~
\min\,(\,A\,,\,B\,)~~~~~~~~~~
\end{eqa*} 
$\mbox{so~ that}~~~~~\,
(\,\nu_m\,-\,1\,)\,(\,\pi\,-\,2\,E\,)\,\ge\,
\min\,(\,A\,,\,B\,)\,+\,
(\,\nu_m\,-\,1\,)\,(\,A\,+\,B\,)~.$ \xmn
$~$Further~~~ $~~\pi\,-\,2\,E~=~
2\,\tan^{-1}(\,c_i\,/\,s\,)~
<~2\,c_i\,/\,s\,.~$ 
It ~follows ~that \xsn
$~~~\play
(\,\nu_m\,-\,1\,)\,a~\ge~
\frac{~(\,1\,-\,\nu_c\,)~}{2\,r_o}~
[~\tan^{-1}(\,r_o\,/\,2\,r\,)~
+~(\,\nu_m\,-\,1\,)
\tan^{-1}\,(\,2\,r_o\,/\,r\,)~]~.$ \xsn
The second inequality in part ~(b)~ of 
the theorem follows from this\,.
\begin{rem}\textsl{
It is clear from the proofs that in both 
the theorems above\,, results 
similar to those of \S\,3 
will hold
if the condition in part (b) 
does not hold\,.}
\end{rem}
\section{Fundamental solutions of ~T\,G\,E}
In this section ~TGE~ will be solved for a 
class of flows\,. It is seen here in a 
partially discreet form as a 
quadratic recursion 
relationship on a sequence of 
smooth functions\,. 
A pair of linearly 
independent solutions is obtained 
in some neighbourhood 
of the critical layer $~~z\,=\,z_c~~$ 
where $~~u(z_c)\,=\,c_r\,,~$ 
in theorem~\tth f when $~~c_i\,
>\,0~~$ is sufficiently small\,. To prove 
the smoothness of the solutions it is assumed 
that $~~[~g\,\beta(z)\,
/\,u^{\,\prime\,2}(z)~]~~$ 
is a constant\,. It is very likely that the 
results hold without this\,. Corollary 
\ccor f~ gives a smooth solution of 
~Rayleigh's equation which vanishes at 
$~~z\,=\,z_c~~$ when $~~c_i\,=\,0\,.~$\xss

Let $~~U~~$ be an open subinterval of \OI. 
Let $~~z_c~~$be a point in $~~U\,.$ ~
Let~$~~u(z)\,,~\beta(z)\,,~$and $~v(z)~~$ 
be $C^\infty-$functions on $~~U~$. For 
$~~z~~$ in $~~U~~$ let \xnl
$~~w(z)\,=\,\exp\,\llb\,\cint \, v(z)\,dz~
\rrb~~$ satisfy ~\xx1\,, then 
\[v^\prime(z)\,+\,v^2(z)\,-\,k^2\,-\,
\frac{\ddu(z)}{~\{\,u(z)\,-\,c\,\}~}\,+\,\frac
{g\,\beta(z)}{~\{\,u(z)\,-\,c\,\}^2~}~=~0~.\]
For $~~n\,\ge\,0~~$ let  $~~a_n(z)~~$ be~ 
$C^\infty-$functions~ on 
$~~U~~$ ~and ~let\xsn
$ ~~v(z)\,=\,
\displaystyle{\frac{\dupm(z)}{~\uc~}\,+\,
\sum_{n=0}^{\infty}~a_n(z)\,Y^n(z)}\,,~$ 
where $~~Y(z)\,=\,\displaystyle
\frac{~u(z)\,-\,c~}{\du(z)}\,.~$ \xsn
Suppose these series converge uniformly 
on some open subset $~~J~~$ of $~~U\,.$ 
Substituting these values of $~~v(z)~~$ 
in turn\,, in the equation above\,, one gets \xmn
$\displaystyle{
\frac{~-\,\ddump(z)\,+\,2\,\dupm(z)
\,a_0~}{~\uc~}
\,+\,\llb\,\llm\,1+\frac{~2\,\dupm(z)~}{\du(z)}
~\rrm\,a_1 +a_o^\prime 
+a_0^2\,-\,k^2~\rrb ~+~}$  \xsn
$\displaystyle{\sum_{n=1}^\infty 
\llb\llm n+1+\frac{\,2\,\dupm(z)~}{\du(z)}
\rrm\,a_{n+1} +a_n^{\,\prime} 
-n\,a_n\, \frac{~\ddu(z)~}{~\du(z)~}+
\sum_{j=0}^n\,a_j\;a_{n-j}\,\rrb Y^n=\,o }\,.$\xsn
This proves part (a) of the following :
\begin{theorem}
\label {thf}
Let ~\ub~ be as above\,. Let 
$~~\beta(z)\,,~\Delta(z)~\ge~0~$ and \xsl
$~~\du(z)~>~0~~$ 
for every $~~z~~$ 
in $~~U\,.$ Let ~\kc~ be ~given~ 
and $~~u(z_c)~=~c_r\,.$
\[~Let~~~~~~~~~
w_1(z)\,=~\exp\,\llb~\cint\,\llm~\frac
{\dup(z)}{~\uc~}\,+\,\sum_{n=0}^\infty\,
a_n(z)\,Y^n(z)~\rrm\,dz\,\rrb~~~ \]
\[and~~~~~~~~
w_2(z)\,=~\exp\,\llb\cint\,\llm~\frac
{\dum(z)}{~\uc~}\,+\,\sum_{n=0}^\infty\,
b_n(z)\,Y^n(z)~\rrm\,dz\,\rrb~~~ \]   
$\,$where $~~~~~\displaystyle
a_0(z)\,=\,\frac{\ddum(z)}{~2\dup(z)~}\,,~~
a_1(z)\,=\,\frac{~\llm\,k^2\,-\,
a_0^\prime(z)\,-\,a_0^2(z)\,\rrm~} 
{1\,+\,2\,\{\,\dup(z)\,/\,\du(z)\,\}}~~$~~~and  
\[ ~a_{\,n+1}(z)\,=\,\frac{1}
{~(n+1)\,+\,2\,\{\,\dup\,/\,\du\,\}~} \llb\,
-a_{\,n}^{\,\prime}+\,n\,a_{\,n}\,
\llm\,\frac{\ddu}
{\du}\,\rrm-\sum_{j=0}^{n}\,a_j~a_{\,n-j}
~\rrb~ \] 
for $~~n\,\ge\,1~~$~ and 
~~ $~b_n(z)~~$ is obtained from 
$~~a_n(z)~~$ on ~interchanging 
$~~u_+(z)~~$ and 
$~~u_-(z)\,.$ \xmn
$(a)~$ Suppose ~the~ series $~~~ \sum~ 
a_n(z)~Y^n(z)~~~[$\,resp.~ $~\sum 
b_n(z)~Y^n(z)~]~~$ 
converges ~uniformly ~in 
~some ~neighbourhood $~~J~~$ of 
$~~z_c\,,~$ then 
~$w_1(z)~~~[$\,resp.~~ $~~w_2(z)~]$ ~~is ~ a  
~smooth ~solution ~of ~TGE~ in $~J\,.$\xmn
$(b)~$ Let ~~Richardson ~number $~~~
\{\,g\,\beta(z)\,/\,
u^{\,\prime\,2}(z)\,\}~~$ be ~a 
~constant ~in 
$~~U\,.$ Let $~~\as~\ge~1\,+\,k^2~~$ 
~be ~a  ~constant\,. ~Suppose ~
for every $~~z~~$ in $~~U\,,$\xsn
$~~~~(1)~~\displaystyle{
\left|~\frac{d^j}{\,dz^j\,}\llm
\frac{\,\ddu(z)\,}{\du(z)}\rrm~
\right|\,\le\,\as^{\,j\,+1}~~~}$ 
~ for ~every $~~j\,\ge\,0~~~$ and \xsn 
$~~~~(2)~~|\,a_0(z)\,|~~\lls \right.$~resp\,. 
$~|\,b_0(z)\,|~\left.\rrs \,\le\,
\as\,,~$ ~then\xmn
for ~sufficiently ~small $~~c_i\,>\,0\,,$ 
$~~\sum~ 
a_n(z)~Y^n(z)~~~~[$~resp.
$~\sum b_n(z)~Y^n(z)~]$\xsl
converges ~absolutely ~and ~uniformly ~in 
~some ~neighbourhood $~~J~~$ of $~~z_c\,.$ 
\end{theorem}
\textit{\textbf {Proof\,:}}~~~
Under ~the ~hypothesis ~of ~part (b)\,, 
$~~\dupm(z)~~$ 
are ~constant~ ~multiples of 
$~~\du(z)\,.$ 
It ~is ~then ~clear ~that 
$~~a_n~~$ and $~~b_n~~$ are ~polynomials 
in $~~[~\ddu(z)\,/\,\du(z)~]~~$ and ~its ~
derivatives ~with 
~constant ~coefficients\,.

Let $~~\displaystyle{
t_j(z)~=~\frac{d^j}{\,dz^j\,}\llm
\frac{\,\ddu(z)\,}{\du(z)}\rrm}~~$ 
for $~~j\,\ge\,0\,.~$ Let $~~R\,=\,
\mathbb R\,[\,t_0\,,~t_1\,,~\,.\,.\,.\,]$ 
be the ring of polynomial functions of 
$~~t_0\,,~t_1\,,\,.\,.\,.~~$ with real 
coefficients\,. For a multi--index $~~
m\,=\,(\,m_0\,,~m_1\,,\,.\,.\,.\,)~~$ with only 
finite number of nonzero entries\,,~ let 
$~~deg\,(m)\,=\,
\sum_{j=0}^{\;\infty}~(j+1)~m_j~~$ 
~and~ $~~t^m\,=\,
\prod_{j=0}^{\;\infty}~ t_j^{m_j}\,.$~\xss

For $~~q(z)\,=\,\sum_m ~q_m~t^m~~$ in 
$~~R\,,~$ let 
$~~deg\,(q)\,=\,\play
\max_m~\{\;deg\,(m)\,~|~\,
q_m\,\ne\,0\;\}~~$ and 
$~~\|\,q\,\|_{\,\as}\,=\, 
\sum_m\,|\,q_m\,|~\as^
{\;deg\,(m)}\,.~$
It is easily checked that 
if $~~q\,,~q_1\,,~q_2~\in~R~~$ 
~and~ $~~\lambda~\in~\mathbb R~$, then \xmn
(1)~~$\play\sup_z |\,q(z)\,|~
\le~\|\,q\,\|_{\as}~~;~~~~~~~~~\,~$ 
(2)~~$\|\,q_1\,+\,q_2\,\|_{\as}~\le~
\|\,q_1\,\|_{\as}\,+\,\|\,q_2\,\|_{\as}~~;$\xsn
(3)~~$\|\,q_1\,q_2\,\|_{\as}\,\le\,
\|\,q_1\,\|_{\as}\,\|\,q_2\,\|_{\as}\,$
~~;~~~
(4)~~$\|\,\lambda\,q\,\|_{\as}\,=\,
|\,\lambda\,|~\|\,q\,\|_{\as}\,~$
~~~~~and \xmn
(5)~~$\|\,q^{\,\prime}\,\|_{\,\as}~\le~
\as~\;deg\,(q)\;~\|\,q\,\|_{\as}~\,.$\xmn
The last part follows because on 
differentiating a monomial of degree 
$~~n~~$ one obtains a sum of at most 
$~~n~~$ 
monomials~ of degree $~~(\,n+1\,)\,.$\xss

It is clear from the hypothesis that~ 
$~~a_n\,,b_n~\in\,R~~$ 
for every $~~n~\ge~0~~$ 
and $~~deg\,(a_n)\,
=\,deg\,(b_n)\,=\,(\,n+1\,)\,.$ 
Further $~~\|\,a_0\,\|_{\,\as}~\le~\as~~$ 
and $~~\|\,a_1\,\|_{\,\as}~\le~3\,\as^{\,2}\,.~$ 
We now prove that 
$~~\|\,a_j\,\|_{\,\as}~\le~3^{\,j}\,
\as^{\,j\,+1}~~$ for every $~~j\,\ge\,0\,.$ 
Inductively we assume that this result 
holds for $~~j~<~n\,,$ then \xsn
$\play{~~~~~~~~~~~
\|\,a_n\,\|_{\,\as}~\le~
\frac{\,1\,}{\,n\,}~
\|\,a_{n-1}^{\,\prime}\,\|_{\,\as}~
+~\as~\|\,a_{n-1}\,\|_{\,\as}~+~
\frac{1}{\,n\,}~\play\sum_{j=0}^
{n-1} 3^{\,n-1}~\as^{\,n+1}\,.}~$~ \xsn
$~~~~~~~~~~~~~~~~~~~~~\,~
\le~3^{\,n}~\as^{\,n+1}~\,,~ $ \xsn
so that ~~~~~~~~~
$\play{\limsup_{n\rightarrow \infty} 
\llb~\sup_z ~|\,a_n(z)\,|^
{ \,\textstyle \frac{1}{n} }~\rrb~
\le~3\,\as\,.}$ \xsn
It follows that $~~\sum_{n=0}^\infty \,
a_n(z)\,Y^n~~$ converges absolutely 
and uniformly for $~~z~~$ in $~~U\,,$ ~and~ 
$~~|\,Y\,|\,\le\,r\,,~$ 
~for~ any~ $~~r\,<\,
1\,/\,[~3\,\as~]\,.$ ~

A similar argument 
with $~~a_n~~$ replaced by $~~b_n~~$ 
gives the same result about the convergence 
of $~~\sum_{n=0}^\infty \,
b_n(z)\,Y^n(z)\,.~$ Part (b) now follows 
from the fact that for sufficiently small 
$~~c_i~>~0\,,~$ the set $~~\{~
|\,Y(z)\,|\,\le\,r~\}~~$ is a closed 
neighbourhood of $~~z_c~~$ in \I.
\begin{rem}
\label {rem6}
\textsc{\,(1)}
\textsl{
The function $~u(z)~$ together with 
its two derivatives can be 
uniformly approximated by a 
function $~~q(z)\,=\bint[\, 
\exp\,\{\bint\,
p\,(z)\,dz\,\}\,dz\,]$ 
where $~~p\,(z)~~$ is a 
polynomial in $~z\,.~$ Clearly  
then $~~[\;q^{\,\prime\prime}(z)\,
/\,q^{\,\prime}(z)\;]~~$ together with all 
its derivatives~ is uniformly 
bounded\,.}\xsn
\textsc{~(2)~}
\textsl{
Suppose $~~c_i\,=\,0\,=\,\beta(z_c)\,=\,
\ddup(z_c)~~$ 
and $~~\T(z)~~$ has a smooth extension 
to ~\I, 
then $~~w_2(z)~~$ is a smooth solution 
of ~TGE\,.
}\end{rem}

When $~~\beta\,\equiv\,0\,,~$ we ~have 
$~~u_+(z)\,=\,u(z)\,,$ ~and~ 
$~u_-(z)\,=\,0\,.~$ 
This ~leads ~to 
~the ~regular ~solution ~of ~Rayleigh's 
~equation\,.
\begin{cor} \label                          {corf}
~~
Suppose $~~\beta(z)\,=\,0~$ ~
and~ $~\du(z)\,>\,0~~$  
for every $~~z~~$ in \xsn
$~~U\,.$ ~Let $~~~\displaystyle
a_1(z)\,=\,{~k^2~}/\,{3}~\,,~~$ and~~
for~ $~~n\,\ge\,1~$\xnl
$\play{~~~~
a_{n+1}(z)\,=\,\frac{1}{~n+3~}~\llb\,
-\,a_{n}^{\,\prime}\,+\,n\,a_{n}\,
\lls\,\frac{\ddu}
{\du}\,\rrs\,-\,
\sum_{j=1}^{n-1}\,a_j~a_{\,n-j}
~\rrb\,.~ }$ \xsn 
$~~~~$ Let $~~~~~\play{w(z)\,=~
\uc~\exp\,\llb~\bint\,
\sum_{n=1}^\infty\,
a_n(z)\;Y^n(z)\;dz\,\rrb\,.}~$\xsn
Suppose ~for ~some $~~\as~
\ge~1\,+\,k^2\,/\,3\,,~~$ 
$\displaystyle{
\left|\,\frac{d^j}{\,dz^j\,}\llm
\frac{\,\ddu(z)\,}{\du(z)}\rrm\,
\right|\,\le\,\as^{\,j\,+1}~~}$
~ for ~every $~~j\,\ge\,0~~$ 
and every $~~z~~$ in $~~U\,.$ 
Let $~~c_i~\ge~0~~$ be ~
sufficiently
small~ then the series 
$~~\sum_{n=0}^\infty \,
a_n(z)\,Y^n(z)~~$ converges absolutely 
and uniformly ~
in ~some ~neighbourhood $~~J~~$ of 
$~~z_c~~$ where $~~u(z_c)\,=\,c_r~~$~ and 
$w(z)~~$ is ~a 
smooth ~solution ~
of ~Rayleigh's equation~ in $~~J\,.$
\end{cor}
\subsection{Existence ~of ~unstable ~flows\,.}

So many parallel flows are believed to be 
unstable because they are not seen to persist\,, 
yet very few can be proved to be so\,. 
The only known example of an unstable 
parallel flow 
with rigid and finite boundaries is 
a piecewise linear flow called a ~shear-
layer ~(\,see~\qqdr\,)\,. The difficulty is 
that no general result 
for the existence of solutions of 
~Taylor-Goldstein boundary value 
problem is known\,. A good beginning in 
this direction has been made~ 
(\,see \qqfh~; \qqfr\,)~ but much 
remains to be done here\,. An existence theorem 
will provide deep insight into the subject\,. We~ 
expect~:\xmn
(1)~~Let ~\ub~ be a flow satisfying 
condition ~(A)~ and let $~~\T(z)~~$ 
be bounded\,. Suppose for every $~~k^2\,\le\,
|\,\T\,|_{\,\max}\,,~$ every $~~c~~$ 
inside ~Howard's semicircle and every 
piecewise$-C^2~$ function $~~f(z)~~$ 
satisfying $~~Im\,[\,f^{\,2}\,]\,<\,0~~$ 
everywhere\,, the condition~: $~~
Im\llb\,ff^{\prime\prime}\,+\,
Af^2\,\rrb\,<\,0~~$ for some $~~z\,,~$ 
holds then the flow is unstable\,. \xmn
(2)~~Suppose for some $~~t_o\,<\,t_1\,<\,t_2
~~$ in ~\I,~ $~~\Delta(t_o)\,<\,0\,,~\Delta(t_2)
\,<\,0~~$ 
$~~~~~~~$and $~~\Delta(t_1)\,>\,0~~$ then 
the flow is unstable\,. \xmn
(3)~~If $~~\beta(z)\,<\,0~~$ 
for some $~~z~~$ then the flow 
is unstable\,. \xmn
{\Large\textbf{Acknowledgements\,:}}\xms

I take this opportunity to remember my teachers 
with reverence and gratitude\,. 
It would have been difficult to carry out 
this investigation without the stable 
financial support and unfailing moral support 
from my family\,. I express a deep 
sense of gratitude for 
this gracious gesture\,.

%

%
}}
\end{document}